\documentclass[fleqn,usenatbib]{mnras}

\usepackage{newtxtext,newtxmath}
\usepackage[T1]{fontenc}
\usepackage{ae,aecompl}

\usepackage{graphicx}	% Including figure files
\usepackage{amsmath}	% Advanced maths commands
\usepackage{amssymb}	% Extra maths symbols
\usepackage{hyperref}
\usepackage{tablefootnote}
\usepackage{wasysym}
\usepackage{float}
\title[Ground-based follow-up of TRAPPIST-1 in the near-IR]{Ground-based follow-up observations of TRAPPIST-1 transits in the near-infrared}

\author[A. Burdanov et al.]{
A. Y. Burdanov,$^{1}$\thanks{E-mail: artem.burdanov@uliege.be}
S. M. Lederer$^{2}$, 
M. Gillon$^{1}$, 
L. Delrez$^{3}$, 
E. Ducrot$^{1}$,
J. de Wit$^{4}$,
\newauthor
E. Jehin$^{5}$,
A. H. M. J. Triaud$^{6}$,
C. Lidman$^{7}$,
L. Spitler$^{8,9}$,
B.-O. Demory$^{10}$,
D. Queloz$^{3}$
\newauthor
and V. Van Grootel$^{5}$
\\
$^{1}$Astrobiology Research Unit,  Universit\'e de Li\`ege, All\'ee du 6 Ao\^ut 19C, 4000 Li\`ege, Belgium\\
$^{2}$NASA Johnson Space Center, 2101 NASA Parkway, Houston, TX 77058, USA\\
$^{3}$Cavendish Laboratory, JJ Thomson Avenue, Cambridge, CB3 0H3, UK \\
$^{4}$Department of Earth, Atmospheric and Planetary Sciences, Massachusetts Institute of Technology, 77 Massachusetts Avenue,\\ Cambridge, MA 02139, USA\\
$^{5}$Space Sciences, Technologies and Astrophysics Research (STAR) Institute, Universit\'e de Li\`ege, All\'ee du 6 Ao\^ut 19C, 4000 Li\`ege, Belgium\\
$^{6}$School of Physics \& Astronomy, University of Birmingham, Edgbaston, Birmingham B15 2TT, UK\\
$^{7}$The Research School of Astronomy and Astrophysics, Australian National University, ACT 2601, Australia\\
$^{8}$Research Centre for Astronomy, Astrophysics \& Astrophotonics, Macquarie University, Sydney, NSW 2109, Australia\\
$^{9}$Department of Physics \& Astronomy, Macquarie University, Sydney, NSW 2109, Australia\\
$^{10}$University of Bern, Center for Space and Habitability, Sidlerstrasse 5, CH-3012 Bern, Switzerland\\
}

\date{Accepted 2019 May 14. Received 2019 May 9; in original form 2019 April 3}

\pubyear{2019}

\begin{document}
\label{firstpage}
\pagerange{\pageref{firstpage}--\pageref{lastpage}}
\maketitle

\begin{abstract}
The TRAPPIST-1 planetary system is a favorable target for the atmospheric characterization of temperate earth-sized exoplanets by means of transmission spectroscopy with the forthcoming \textit{James Webb Space Telescope (JWST)}. A possible obstacle to this technique could come from the photospheric heterogeneity of the host star that could affect planetary signatures in the transit transmission spectra. To constrain further this possibility, we gathered an extensive photometric data set of 25 TRAPPIST-1 transits observed in the near-IR J band (1.2~$\mu$m) with the \textit{UKIRT} and the \textit{AAT}, and in the NB2090 band (2.1~$\mu$m) with the \textit{VLT} during the period 2015-2018. In our analysis of these data, we used a special strategy aiming to ensure uniformity in our measurements and robustness in our conclusions. We reach a photometric precision of $\sim0.003$ (RMS of the residuals), and we detect no significant temporal variations of transit depths of TRAPPIST-1 b, c, e, and g over the period of three years. The few transit depths measured for planets d and f hint towards some level of variability, but more measurements will be required for confirmation. Our depth measurements for planets b and c disagree with the stellar contamination spectra originating from the possible existence of bright spots of temperature 4500 K. We report updated transmission spectra for the six inner planets of the system which are globally flat for planets b and g and some structures are seen for planets c, d, e, and f.
\end{abstract}

% The near-IR spectral range encompasses the peak of the spectral energy distribution of TRAPPIST-1 and our observations could bring strong constraints on the photospheric properties of the host star and on the impact of stellar contamination to be expected for upcoming \textit{JWST} observations of the planets.


%%%%%%%%%%%%%%%%%%%%%%%%%%%%%%%%%%%%%%%%%%%%%%%%%%

%%%%%%%%%%%%%%%%% BODY OF PAPER %%%%%%%%%%%%%%%%%%

\section{Introduction}

TRAPPIST-1 is a $\sim0.09 M_\odot$ ultracool dwarf star 12.4~pc away in the Aquarius constellation \citep{2000AJ....120.1085G,2018ApJ...853...30V}. Photometric monitoring revealed that it hosts a compact resonant system composed of seven transiting Earth-sized planets \citep{Gillon2016,Gillon2017,Luger2017}. Dynamical modeling of this system based on timing variations of its planets' transits resulted in strong constraints on the planetary masses which, when combined with the radii measured from transits, point toward rocky compositions with sizable volatile contents \citep{Grimm2018}. With stellar irradiations ranging from 0.1 to 4 times solar, these seven planets can be qualified as "temperate", and three of them (planets e, f, and g) orbit within the "habitable zone" of the host star \citep{Gillon2017}. 

Due to the combination of relatively large infrared brightness ($J \sim 11.4$, $H \sim 10.7$, $K \sim 10.3$) and the small size ($\sim 0.12 R_\odot$) of their host star, the TRAPPIST-1 planets are especially favorable targets for the detailed exploration of their atmospheres with the future ground-based extremely large telescopes and \textit{James Webb Space Telescope (JWST)} \citep{2014ApJ...781...54R,Barstow2016,Gillon2017,Morley2017,2018ApJ...867...76L}. First reconnaissance observations with the \textit{Hubble Space Telescope (HST)} by \cite{deWit2016,deWit2018} showed no hint of clear $\mathrm{H/He}$ dominated atmospheres for the six inner planets. Further studies reassessed these findings: \cite{2018AJ....156..252M} supported the presence of secondary volatile-rich atmospheres for planets d, e and f using revised masses (while not entirely ruling out a cloud-free $\mathrm{H}$-rich atmosphere for TRAPPIST-1\,e and f) and \cite{2019AJ....157...11W} ruled out a clear solar $\mathrm{H_2/He}$ dominated atmosphere for TRAPPIST-1\,g.

A possible obstacle on the way for a proper and robust characterization of the atmospheres of the TRAPPIST-1 planets was presented by \cite{Apai} and \cite{Rackham2018}, who argue that the photospheric heterogeneity of the host star, and specifically spots and faculae, could alter, hide, or even mimic planetary signatures in the transit transmission spectra. Taking into account this possible stellar contamination, and basing their re-analysis on existing \textit{HST} near-IR data, \cite{Zhang2018} predicted considerable changes of the transit depths of TRAPPIST-1 planets with wavelength. Work by \cite{2018AJ....156..218D} disproved these predictions by producing featureless broadband transmission spectra for the planets in the 0.8-4.5~$\mu$m spectral range, and showing an absence of significant temporal variations of the transit depths in the visible. Furthermore, \cite{Delrez2018} showed that the 3.3~d periodic 1\% photometric modulation detected in the \textit{K2} optical data set \citep{Luger2017} is not present in \textit{Spitzer} 4.5~$\mu$m observations, and  that transit depth measurements do not show any hint of  a  significant stellar contamination in this spectral range. The same conclusion was reached by  \cite{2018ApJ...863L..32M} using a "self-contamination" approach based on the measurement of transit egress and ingress durations from the same \textit{Spitzer} data set. 

Based on the presence of a 3.3~d photometric variability in \textit{K2} data and its absence in \textit{Spitzer} mid-IR data, \cite{2018ApJ...857...39M} proposed that the photospheric inhomogeneities of TRAPPIST-1 could be dominated by a few faculae (i.e. bright spots) with characteristic temperatures in excess of $\sim$ 1800~K relative to the effective temperature of the star ($\sim 2500$ K). This is globally consistent with the analysis of \cite{Rackham2018} and \cite{Zhang2018} who inferred whole-disk spot and faculae coverage fractions of $\sim$10 and $\sim50$\% respectively, but with a much lower temperature excess of ($\sim$ 500K) for the faculae. According to \cite{2018AJ....156..218D}, their analyses favor two scenarios of TRAPPIST-1 photosphere: with a prevalence of a several high-latitude cold and large spots, or alternatively, a few hot and small spots.  

Globally, all these studies do not yet present a consistent picture of the photospheric properties of TRAPPSIT-1, even at the observational level. For instance, the 1.1-1.7 $\mu$m combined transmission spectra presented by \cite{2018AJ....156..218D} do not show any significant features, while the spectra obtained by \cite{Zhang2018} from the very same \textit{HST} data set shows a drop around 1.4 $\mu$m attributed by the authors to an inverted water absorption feature caused by stellar contamination. Both studies agree that the \textit{HST} transit depths are globally deeper than those measured at other wavelengths. Nevertheless, as outlined by \cite{2018AJ....156..218D}, the origins of these larger transit depths could be instrumental, as the \textit{HST/WFC3} systematic effects combined with the low-Earth orbit of \textit{HST} make it very difficult to normalize transit light curves as short as those of TRAPPIST-1 planets (see Subsection 3.2 in \citealt{2018AJ....156..218D}). The uncertainties affecting the absolute values of the transits measured by \textit{HST} in the near-IR  is unfortunate, as this spectral range encompasses the peak of the spectral energy distribution of TRAPPIST-1 (see Fig.~3 in \citealt{2018ApJ...857...39M}) and could bring strong constraints on the photospheric properties of TRAPPIST-1 and on the impact of stellar contamination to be expected for upcoming \textit{JWST} observations of the  planets. As Earth's atmosphere is partially transparent in near-IR, it is therefore highly desirable to observe as many transits of  TRAPPIST-1 planets as possible in this spectral range at high precision with ground-based telescopes.

In this paper, we present an extensive photometric data set of 25 TRAPPIST-1 transits gathered in the near-IR J~band (1.2~$\mu$m) with the 3.8-m \textit{United Kingdom Infra-Red Telescope (UKIRT)} and the 3.9-m \textit{Anglo-Australian Telescope (AAT)}, and in the NB2090 band (2.1~$\mu$m) with the ESO \textit{Very Large Telescope (VLT)} during the period 2015-2018. In our analysis of these data, we used a special strategy (see Subsections \ref{phot_ext} and \ref{ind_an}) aiming to ensure uniformity in our measurements and robustness in our conclusions. This special attention is motivated by the inherent complexity of ground-based near-IR data reduction appearing as correlations of deduced transit depths with photometric aperture sizes and comparison stars (for a comprehensive review of this topic see \citealt{2015ApJ...802...28C}). We also report precise timings of each transit that should further constrain the masses of the planets via the transit-timing variations method \citep{Gillon2017,Grimm2018}.

The rest of the paper is divided into four sections. First, we describe our observational data set in Section~\ref{INSTRUMENTS AND OBSERVATIONS}. Section~\ref{DATA REDUCTION AND ANALYSIS} is devoted to the reduction and analysis of these data. We present and discuss our results in Section~\ref{RESULTS AND DISCUSSION}, and we outline our findings in the \hyperref[CONCL]{Conclusions} section.

\section{INSTRUMENTS AND OBSERVATIONS}\label{INSTRUMENTS AND OBSERVATIONS}

\begin{figure}
    \centering
    \includegraphics[width=\columnwidth]{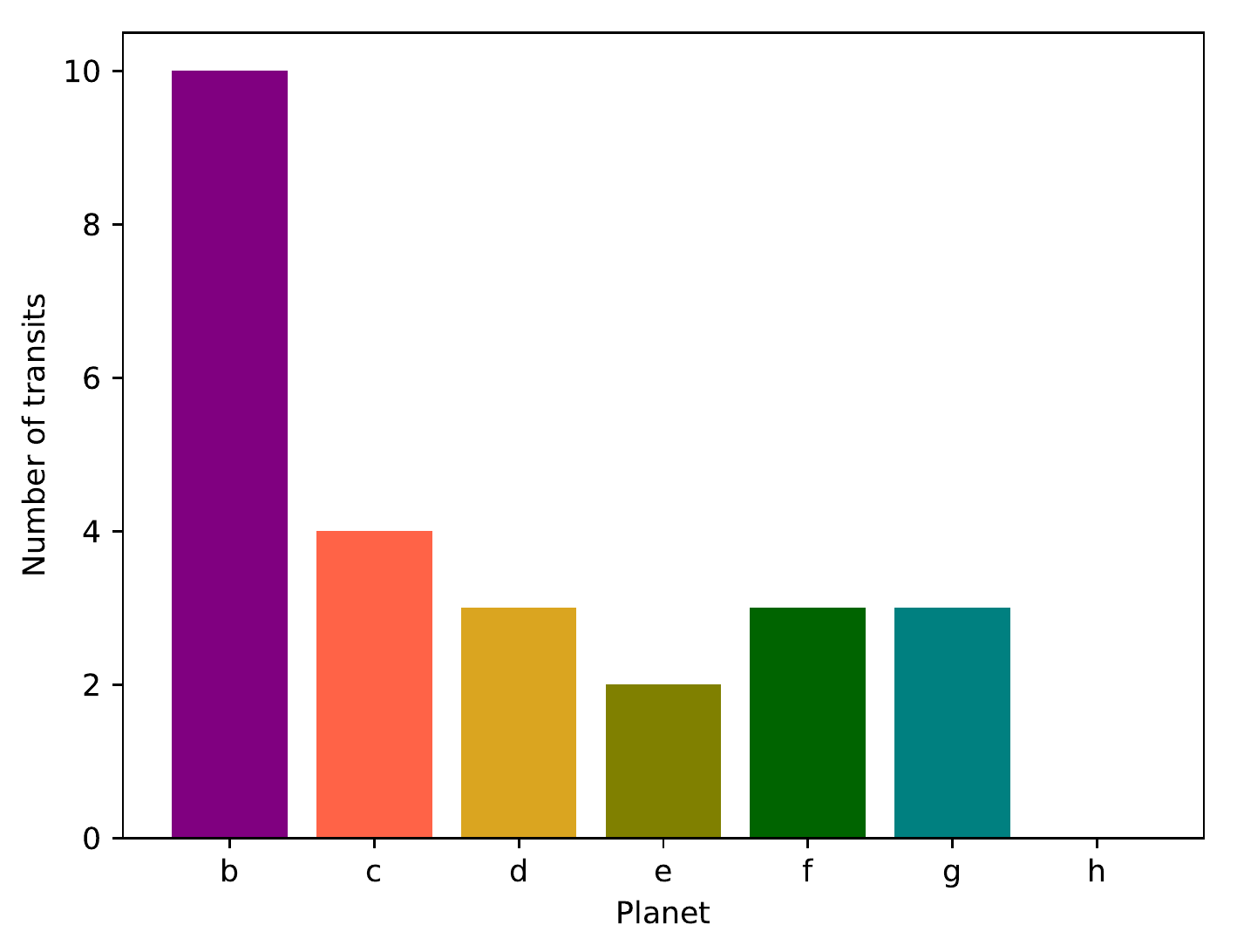}
    \caption{Distribution of the TRAPPIST-1 planet transits analyzed in this work.}
    \label{histo}
\end{figure}

We acquired more than 30 transits of TRAPPIST-1 planets with \textit{UKIRT, VLT}, and \textit{AAT} in the period from December 2015 to July 2018 through different observing programs (see \hyperref[Ack]{Acknowledgements} section for a full list of programs). For our analysis, we considered only 25 transits which satisfied the following two conditions: (1) isolated transits, i.e. not blended with the transit of another planet, and (2)  taken under relatively good observing conditions, meaning transparency variations in the Earth's atmosphere were less than 30\%. The distribution of these 25 transits among the different TRAPPIST-1 planets is shown in Fig.~\ref{histo}.

All transits were observed in "staring mode", i.e. without dithering of the telescope. However, before and after the scientific sequence, sky flat images were acquired with dithering to construct a proper master sky flat image. An observing log is presented in Table~\ref{tab:sets}, and we outline instrument-specific information below.

\subsection{UKIRT/WFCAM}

The WFCAM near-IR imager of the \textit{UKIRT} 3.8-meter telescope located on the summit of Mauna Kea was used to observe transits of TRAPPIST-1\,b, c, d, e, f and g in the broad-band J filter\footnote{\url{http://casu.ast.cam.ac.uk/surveys-projects/wfcam/technical/filter-set}}. Those observations occurred between 2015 and 2018. The WFCAM imager consists of four $2048\times2048$ Rockwell Hawaii-II detectors with a field of view (FoV) of $13.65\times13.65$~$\mathrm{arcmin^2}$ each and image scale of 0.4~arcsec~pixel$^{-1}$ \citep{2007A&A...467..777C}. TRAPPIST-1 was placed in  quadrant 3 (array ID number \#76) as this is the cleanest of the four detectors. 

Observations were made in Correlated Double Sampling (CDS) mode which is the default read mode used for all broad-band observations. These observations comprised co-adding five x 1-second exposures throughout the undithered scientific sequence. The sequence was extended long enough - 1 to 2.5 hours on each side of the transit, depending upon how well known the orbital period of the target planet's transit was known - to ensure that a pre- and post-transit, star-only baseline was acquired. For example, a 1~h transit of planet f, plus 1.5~h to capture the baseline out-of-transit prior to and following the transit, resulted in a 4~h total undithered scientific sequence. The dithered sky flat sequence before and after each full scientific transit sequence was composed of capturing five x 1-second co-added exposures throughout the 2.5~min sequence.

\subsection{VLT/HAWK-I}\label{subsec:vlt}

We used the  HAWK-I cryogenic wide-field imager installed on Unit Telescope 4 (Yepun) of the ESO \textit{VLT} at Paranal observatory to observe transits of TRAPPIST-1\,b and c in 2015 and 2017. The HAWK-I imager is composed of four $2048\times2048$ Hawaii 2RG chips \citep{2011Msngr.144....9S}. Each chip provides an image scale of 0.106~arcsec~pixel$^{-1}$ resulting in a $217\times217$~$\mathrm{arcsec^2}$ FoV. TRAPPIST-1 was placed in the corner of the quadrant Q3 (chip \#79) to allow three additional stars to be simultaneously imaged on the chip for use as comparison stars. Non-Destructive Read (NDR) was used with a 3 second exposure time and 12 sub-integrations. All transits were observed in narrow-band filter NB2090 which has a central wavelength of 2.095~$\micron$ and width of 0.020~$\micron$. The  small width of this filter minimizes the effect of differential extinction, while the combination of its central wavelength and bandwidth eliminates large absorption and emission bands present in the K band\footnote{\url{https://www.eso.org/sci/facilities/paranal/instruments/hawki/inst.html}}.  

\subsection{AAT/IRIS2}
We  observed two transits of TRAPPIST-1\,b with the IRIS2 IR-imager installed on the \textit{AAT} 4-m telescope at the Siding Spring Observatory. IRIS2 IR-imager consists of one $1024\times1024$ Rockwell Hawaii-II detector which has a FoV of $7.7\times7.7$~$\mathrm{arcmin^2}$ and a pixel scale of 0.45~arcsec~pixel$^{-1}$ \citep{2004SPIE.5492..998T}. We used 9~second exposure times and observations were done in the J band. The telescope was pointed in such a way as to prevent TRAPPIST-1 and comparison stars from falling in the upper right quadrant of IRIS2, which had excessive noise.       
\par
\begin{table*}
\small
\caption{Observing log: observations are grouped by planets and sorted chronologically in each group.}
\noindent\makebox[\textwidth][c]{
\begin{minipage}{1\textwidth}
\begin{tabular}{ccccccccc}
\hline
ID & Date of start & Planet & Telescope/Instrum. & Filter & Duration & Exposure time & Sky transparency\footnote{According to the ESO definitions (\url{https://www.eso.org/sci/observing/phase2/ObsConditions.html})} & Remarks\\
\phantom{text} & of the night & \phantom{text} & \phantom{text} & \phantom{text} & \phantom{text} & (N$\times$EXP)\footnote{N is the number of exposures and EXP is the individual exposure times} & \phantom{text}\\
\hline
%---------------------------------------b---------------------------------------------------------------------
1 & 2015 Nov 07 & b & VLT/HAWK-I & NB2090 & $\sim4.0$~h & $12\times$3.0~s & Clear & -\\
2 & 2015 Dec 05 & b & UKIRT/WFCAM & J & $\sim3.5$~h & $3\times$2.0~s & Photometric & -\\
3 & 2015 Dec 08 & b & UKIRT/WFCAM & J & $\sim3.6$~h & $5\times$1.0~s & Photometric & Partial\\
4 & 2016 Jul 29 & b & UKIRT/WFCAM & J & $\sim 5.6 $~h & $5\times$1.0~s & Photometric & -\\
5 & 2016 Aug 01 & b & UKIRT/WFCAM & J & $\sim 3.2 $~h & $5\times$1.0~s & Photometric & -\\
6 & 2016 Oct 17 & b & AAT/IRIS2 & J & $\sim 3.7 $~h & $1\times$9.0~s & Photometric & 9~min gap\\
\phantom{text} & \phantom{text} & \phantom{text} & \phantom{text} & \phantom{text} & \phantom{text} & \phantom{text} & \phantom{text} & during bottom\\
7 & 2016 Oct 20 & b & AAT/IRIS2 & J & $\sim 3.4 $~h & $1\times$9.0~s & Photometric & Airmass \textgreater 2\\
\phantom{text} & \phantom{text} & \phantom{text} & \phantom{text} & \phantom{text} & \phantom{text} & \phantom{text} & \phantom{text} &  from 7682.13\\
8 & 2017 Jul 07 & b & UKIRT/WFCAM & J & $\sim 2.0 $~h & $5\times$1.0~s & Photometric & Short OOT\footnote{out-of-transit observations},\\
\phantom{text} & \phantom{text} & \phantom{text} & \phantom{text} & \phantom{text} & \phantom{text} & \phantom{text} & \phantom{text} &  gap during egress\\
9 & 2017 Nov 30 & b & VLT/HAWK-I & NB2090 & $\sim2.8$~h & $12\times$3.0~s & Photometric & -\\
10 & 2018 Jul 06 & b & UKIRT/WFCAM & J & $\sim 1.8 $~h & $5\times$1.0~s & Photometric & Short OOT, gap\\
\phantom{text} & \phantom{text} & \phantom{text} & \phantom{text} & \phantom{text} & \phantom{text} & \phantom{text} & \phantom{text} &  before engress\\
\\
%---------------------------------------c---------------------------------------------------------------------
11 & 2015 Dec 06 & c & UKIRT/WFCAM & J & $\sim3.8$~h & $5\times$1.0~s & Photometric & -\\
12 & 2016 Jul 18 & c & UKIRT/WFCAM & J & $\sim4.9$~h & $5\times$1.0~s & Photometric & -\\
13 & 2016 Jul 30 & c & UKIRT/WFCAM & J & $\sim3.3$~h & $5\times$1.0~s & Photometric & -\\
14 & 2017 Nov 21 & c & VLT/HAWK-I & NB2090 & $\sim3.1$~h & $12\times$3.1~s & Photometric & -\\
\\
%---------------------------------------d---------------------------------------------------------------------
15 & 2017 Aug 16 & d & UKIRT/WFCAM & J & $\sim3.0$~h & $5\times$1.0~s & Photometric & Short gaps during\\ 
\phantom{text} & \phantom{text} & \phantom{text} & \phantom{text} & \phantom{text} & \phantom{text} & \phantom{text} & \phantom{text} & egress and bottom\\
16 & 2017 Aug 20 & d & UKIRT/WFCAM & J & $\sim3.0$~h & $5\times$1.0~s & Photometric & Short gap during\\ 
\phantom{text} & \phantom{text} & \phantom{text} & \phantom{text} & \phantom{text} & \phantom{text} & \phantom{text} & \phantom{text} & bottom\\
17 & 2017 Oct 28 & d & UKIRT/WFCAM & J & $\sim2.9$~h & $5\times$1.0~s & Photometric & -\\
\\
%---------------------------------------e---------------------------------------------------------------------
18 & 2017 Sep 11 & e & UKIRT/WFCAM & J & $\sim2.4$~h & $5\times$1.0~s & Photometric & Short gap\\ 
\phantom{text} & \phantom{text} & \phantom{text} & \phantom{text} & \phantom{text} & \phantom{text} & \phantom{text} & \phantom{text} & after egress\\
19 & 2018 Jul 13 & e & UKIRT/WFCAM & J & $\sim2.9$~h & $5\times$1.0~s & Photometric & -\\
\\
%---------------------------------------f---------------------------------------------------------------------
20 & 2017 Jul 31 & f & UKIRT/WFCAM & J & $\sim3.8$~h & $5\times$1.0~s & Thin cirrus & -\\
21 & 2017 Sep 15 & f & UKIRT/WFCAM & J & $\sim3.2$~h & $5\times$1.0~s & Photometric & Short gap\\ 
\phantom{text} & \phantom{text} & \phantom{text} & \phantom{text} & \phantom{text} & \phantom{text} & \phantom{text} & \phantom{text} & before ingress\\
22 & 2017 Oct 22 & f & UKIRT/WFCAM & J & $\sim3.9$~h & $5\times$1.0~s & Thin cirrus & -\\ 
\\
%---------------------------------------g---------------------------------------------------------------------
23 & 2017 Sep 02 & g & UKIRT/WFCAM & J & $\sim5.0$~h & $5\times$1.0~s & Photometric & -\\
24 & 2017 Dec 10 & g & UKIRT/WFCAM & J & $\sim3.5$~h & $5\times$1.0~s & Photometric & -\\
25 & 2018 Jun 13 & g & UKIRT/WFCAM & J & $\sim2.6$~h & $5\times$1.0~s & Photometric & Short OOT, gap\\
\phantom{text} & \phantom{text} & \phantom{text} & \phantom{text} & \phantom{text} & \phantom{text} & \phantom{text} & \phantom{text} &  during ingress\\
\hline
\label{tab:sets}
\end{tabular}
\end{minipage}}
\end{table*}

\section{DATA REDUCTION AND ANALYSIS}\label{DATA REDUCTION AND ANALYSIS}

In this section, we describe the entire data handling process which consisted of a preliminary reduction of the images, photometric extraction of the stellar fluxes, performing differential photometry, and deducing transit parameters with the use of an adaptive Markov Chain Monte Carlo (MCMC) code \citep{2012A&A...542A...4G,2014A&A...563A..21G}. Each transit was first analysed individually to obtain its parameters and to search for possible temporal variability (subsection~\ref{ind_an}). For each planet, global MCMC analyses of all the transits were then performed (section~\ref{glob_an}), although separately for each filter.  

\subsection{Image calibrations}

\subsubsection{UKIRT/WFCAM}
All images obtained with the \textit{UKIRT} telescope were pre-processed by the Cambridge Astronomy Survey Unit (CASU) and then were downloaded from the WFCAM Science Archive (WSA). Reduction steps completed by CASU include: dark-correction, flat-fielding, gain-correction and decurtaining (a specific correction for WFCAM\footnote{\url{http://casu.ast.cam.ac.uk/surveys-projects/wfcam/technical/decurtaining}} data). No linearity correction was applied as the system is linear to <\,1\% up to the saturation regime\footnote{\url{http://casu.ast.cam.ac.uk/surveys-projects/wfcam/technical/linearity}} ($\sim\!40\,000$ counts/pixel) and maximum counts on TRAPPIST-1, which was the brightest star in the FoV, never exceeded $15\,000$ counts. Then the values of the bad pixels (deviating by $3\sigma$ when classified as background sky values and $40\sigma$ for stars, where $\sigma$ is sky background noise) were replaced with the median values of the neighbouring pixels. 

\subsubsection{VLT/HAWK-I}
Raw scientific images and corresponding processed calibration images were downloaded from the ESO archive. Before preliminary data reduction, sub-images from quadrant Q3 were extracted, and all subsequent steps were performed only for quadrant Q3. We used only one quadrant as using reference stars located on other chips reduces photometric precision \citep{2013A&A...552A...2L}. This also minimized computational time for data reduction. Processed calibration files from the ESO archive were used for dark and flat field corrections with the PyRAF/ccdproc\footnote{\url{http://stsdas.stsci.edu/cgi-bin/gethelp.cgi?ccdproc}} module. Then we followed the same procedure outlined for the \textit{UKIRT} data to deal with the bad pixels. As with \textit{UKIRT} data, no linearity correction was applied because the detector is linear to <\,1\% below $30\,000$ counts and maximum counts on TRAPPIST-1 were below $20\,000$.   

\subsubsection{AAT/IRIS2}
All the images from \textit{AAT} were treated similarly to the data from \textit{VLT} except for the manual creation of the master dark and flat images.

\subsection{Photometric extraction}\label{phot_ext}

After the preliminary reduction steps described above, we applied the subsequent procedures to the data from all the telescopes.

All images of each observing run were aligned with respect to the first image of the run (typical X and Y shifts were less than 1-2 pixels). Then we created a median stacked image from all the aligned images to run a star identification algorithm. Positions of the stars on the stacked image were identified with intensity-weighted centroids. Then we extracted fluxes of the stars on each image with eight different circular apertures, sky buffers, and sky annuli using DAOPHOT \citep{Tody1986}. Aperture sizes were defined as $1, 1.5, 2, 2.5, 3, 4, 5, 6 \,\times\,$FWHM, where FWHM is the mean full width at half maximum of the star's point spread function (PSF) in the image. For each star, the sky background was measured in an annulus beyond the stellar aperture using the median sky fitting algorithm implemented in PyRAF/DAOPHOT. Radius of the annulus was defined as $3\times$ FWHM, and its width as $5\times$ FWHM.

At that stage, differential photometry was carried out using a custom-made code: for all apertures, a given target star $T$ and a list of comparison stars ($C1, C2,$, etc...), we computed the ratio $F_{T}/(F_{C1}+F_{C2}+...)$ for each image, where $F$ is the flux of a star corrected for the sky background. We derived photometric uncertainties with the use of a "CCD equation" \citep{2006hca..book.....H} taking into account dark current, read-out noise, stellar scintillation and sky Poisson noise.

At first, we followed the approach to differential photometry and transit light curve analysis that is generally used for ground-based transit photometry obtained in the visible \citep{2018AJ....156..218D}. The best aperture size and combination of comparison stars were selected on the basis of the minimization of the out-of-transit (OOT) scatter of the light curve. Similar to other scientists dealing with near-IR ground-based observations, we found out that this approach is not optimal as it appears to induce correlations of transit depths deduced from the MCMC analysis with aperture size and with the used comparison stars (see \citealt{2015ApJ...802...28C} and \citealt{2018A&A...615A..86C}). While the exact sources of these correlations are unknown, we suspect that more likely they are coming from a combination of these factors: brightness and color differences of the TRAPPIST-1 and reference stars (the target star is the brightest and the reddest star in the FoV while reference stars are fainter and much bluer); spatial separation of the target and reference stars on the sky and on the detector, which make effects of the detector systematics and of the Earth's atmosphere display themselves differently for the target and reference stars. 

We found that different combinations of the reference stars and aperture sizes can be equally good in terms of the OOT scatter, but give significantly different transit depths for a given observing run. Thus, if such an approach is applied to the entire TRAPPIST-1 data set, then one can not truly discriminate temporal variations of the transit depths caused by this near-IR photometry effect, or caused by physical stellar contamination coming from the star, e.g., from unocculted star spots, or caused by some other astrophysical process. Therefore, selection of the photometric aperture size and suitable comparison stars is a non-trivial and important step. In the next section, we describe how we rigorously approached our analysis to yield credible and robust transit depths, as well as transit timings, durations and impact parameters.

\par
\begin{table*}
\small
\caption{Details of TRAPPIST-1 transit light curve analyses. Baseline function is a polynomial of time ($t$), mean FWHM of the star's PSF in the image ($fwhm$), position of the star on the CCD ($xy$), airmass ($A$) and/or sky background ($sky$). The epoch is calculated using the transit ephemeris reported in \protect\cite{Delrez2018}}.
%\noindent\makebox[\textwidth][c]{
%\begin{minipage}{\textwidth}
\begin{tabular}{ccccccccc}
\hline
ID & Planet & Telescope/Instrum. & Filter & Number of points & Epoch & Baseline\\
% $\beta_w$ & $\beta$  & 
\hline
%---------------------------------------b---------------------------------------------------------------------
1 & b & VLT/HAWK-I & NB2090 & 195 & 8 & $t^2+fwhm^2+xy^2$ &\\
2 & b & UKIRT/WFCAM & J & 880 & 26 & $t^3+sky^2+xy^1$ &\\
3 & b & UKIRT/WFCAM & J & 1107 & 28 & $t^1+fwhm^1+sky^3+xy^1$ &\\
4 & b & UKIRT/WFCAM & J & 1558 & 183 & $t^2+fwhm^1+xy^2$ &\\
5 & b & UKIRT/WFCAM & J & 898 & 185 & $t^2+fwhm^3+xy^1$ &\\
6 & b & AAT/IRIS2 & J & 1211 & 236 & $t^2+fwhm^3+xy^3$ &\\
7 & b & AAT/IRIS2 & J & 1916 & 238 & $t^2+fwhm^3+sky^1+xy^3$ &\\
8 & b & UKIRT/WFCAM & J & 584 & 410 & $t^2+fwhm^4+sky^1+xy^1$ &\\
9 & b & VLT/HAWK-I & NB2090 & 152 & 507 & $t^3+fwhm^2+sky^1+xy^2$ &\\
10 & b & UKIRT/WFCAM & J & 516 & 651 & $t^1+fwhm^1+xy^1+sky^2$ &\\
\\
%-------------------------c---------------------------------------------------------------------
11 & c & UKIRT/WFCAM & J & 1148 & 33 & $t^2+fwhm^2+xy^1+sky^1$ &\\
12 & c & UKIRT/WFCAM & J & 1451 & 126 & $t^2+fwhm^2+sky^2+xy^2$ &\\
13 & c & UKIRT/WFCAM & J & 1003 & 131 & $t^4+fwhm^2+xy^1$ &\\
14 & c & VLT/HAWK-I & NB2090 & 169 & 329 & $t^2+fwhm^2+sky^2+xy^4$ &\\
\\
%-------------------------d---------------------------------------------------------------------
15 & d & UKIRT/WFCAM & J & 832 & 77 & $t^2+fwhm^2+xy^2$ &\\
16 & d & UKIRT/WFCAM & J & 866 & 78 & $t^2+A^2+fwhm^1+xy^1+sky^1$ &\\
17 & d & UKIRT/WFCAM & J & 886 & 95 & $t^2+fwhm^2+xy^1+sky^2$ &\\
\\
%-------------------------e---------------------------------------------------------------------
18 & e & UKIRT/WFCAM & J & 697 & 57 & $t^2+fwhm^2+xy^2+sky^1$ &\\
19 & e & UKIRT/WFCAM & J & 832 & 107 & $t^2+fwhm^3+xy^2$ &\\
\\
%-------------------------f---------------------------------------------------------------------
20 & f & UKIRT/WFCAM & J & 1123 & 32 & $t^3+fwhm^3+xy^2$ &\\
21 & f & UKIRT/WFCAM & J & 822 & 37 & $fwhm^1+xy^2$ &\\ 
22 & f & UKIRT/WFCAM & J & 950 & 41 & $t^2+fwhm^2+xy^2+sky^2$ &\\ 
\\
%-------------------------g---------------------------------------------------------------------
23 & g & UKIRT/WFCAM & J & 1468 & 27 & $t^2+fwhm^2+xy^2$ &\\
24 & g & UKIRT/WFCAM & J & 1057 & 35 & $t^2+fwhm^4+xy^2$ &\\
25 & g & UKIRT/WFCAM & J & 715 & 50 & $A^2+fwhm^3+xy^2$ &\\
\hline
\end{tabular}
%\end{minipage}}
\label{tab:baselines}
\end{table*}
\par

\subsection{Individual analysis}\label{ind_an}

One way to robustly determine output metrics is to begin by producing a large set of differential light curves for each transit, based on different combinations of apertures and comparison stars. Each light curve within this set is then analyzed with the MCMC code and the metric RMS$\,\times\,\beta_r^2$ (proposed by \citealt{2015ApJ...802...28C}) calculated for each. Only those combinations whose output metric meet the criteria of RMS$\,\times\,\beta_r^2 \,<\,1.15 \,\times\, [\mathrm{RMS}\,\times\,\beta_r^2]_{min}$ (less than 15\% above the minimum metric) are selected.

Here RMS is the root mean square of the photometric residuals after the subtraction of the transit best-fit model and $\beta_r$ is a quantitative factor used to assess the amount of red (correlated) noise in our time-series \citep{2008ApJ...683.1076W}. After this selection is done, posterior probability distribution functions (PDFs) from the MCMC runs of the selected light curves are combined for each transit parameter. Use of this metric allows us to find a proper balance between small aperture radii, which tend to produce the smallest RMS, and large aperture radii, which minimizes the time-correlated noise ($\beta_r \sim 1$).

For our MCMC analyses, we used the quadratic limb-darkening (LD) law coefficients $u_1$ and $u_2$ with a normal prior distributions. LD coefficients values and corresponding uncertainties were interpolated from the paper by \cite{2012A&A...546A..14C}. Respective values for J~band were $u_1=0.19\pm0.04$ and $u_2=0.50\pm0.12$, and for the NB2090 band -- $u_1=0.23\pm0.05$ and $u_2=0.33\pm0.08$. 

Each transiting light curve was modeled using the model by \cite{2002ApJ...580L.171M} multiplied by a baseline polynomial  model aiming to  account for the correlation of the measured fluxes with the variations of external parameters such as the x- and y-drift of the stars on a chip due to imperfect telescope tracking, FWHM, time, airmass, sky background values, etc.  For each observing run, we first selected a baseline model by running a relatively short chain of MCMC with 10\,000 steps with different combinations of external parameters on a limited set of light curves which were, in their turn, obtained with varying combinations of comparison stars and aperture sizes. A model giving the minimum Bayesian Information Criterion (BIC, \citealt{1978AnSta...6..461S}) was selected, and this model worked efficiently for all the light curves within one observing night. Typically, accounting for the drift of the positions of the stars, for sky background variations, and FWHM changes decreased the BIC significantly. Selected baseline functions for each observing run are presented in Table~\ref{tab:baselines}. 

For each light curve of every observing run, we allowed the following parameters to vary (jump parameters): the mid-transit time $T_0$; the ratio of the planet's and star's areas $(R_p/R_\star)^2$, where the planetary radius is $R_p$ and the stellar radius is $R_\star$; the transit width $W$ (duration from the first to last contact); the impact parameter $b^\prime = a\,cos\,i_p/R_\star$ assuming circular orbit, where $a$ is the semi-major axis and $i_p$ is the orbital inclination; the combinations $c_1 =2u_1 +u_2$ and $c_2 = u_1 - 2u_2$ where  $u_1$ and $u_2$ are the quadratic limb-darkening coefficients. The orbital period for each planet was kept fixed to the values presented in \cite{Delrez2018}. Uniform non-informative prior distributions were assumed for all jump parameters. TRAPPIST-1's effective temperature $T_{\mathrm{eff}}=2516\pm41~K$, its mass $M_{\star}=0.089\pm0.006~M_{\astrosun}$ and its metallicity \lbrack Fe/H\rbrack = $0.04\pm0.08$ were applied from \cite{2018ApJ...853...30V} with normal prior distributions.

For every transit observation, we formed differential light curves using all possible combinations of the comparison stars and aperture sizes. Typically, there were 5-6 suitable comparison stars, and with eight apertures, it produced 400-500 individual light curves for a given transit. However, in the case of \textit{VLT} data, we had just two suitable reference stars. For each light curve, we ran MCMC with $40\,000$ steps to estimate a  best-fit value for the metric RMS$\,\times\,\beta_r^2$, to select light curves and to compute correction factors CF to rescale photometric error bars (see \citealt{2012A&A...542A...4G} for the details). Then for each best light curve, an MCMC model with two chains of $100\,000$ steps were executed with correction factors. We controlled the convergence of the chains by applying the statistical test of \cite{1992StaSc...7..457G} and checking that it was less than 1.11 for each jump parameter. Posterior PDFs derived from the analysis of these selected light curves were combined afterward for each transit parameter. Our MCMC simulations were parallelized and were done with the use of the Consortium des {\'E}quipements de Calcul Intensif (C{\'E}CI) computing center.

\begin{figure}
    \centering
    \includegraphics[width=\columnwidth]{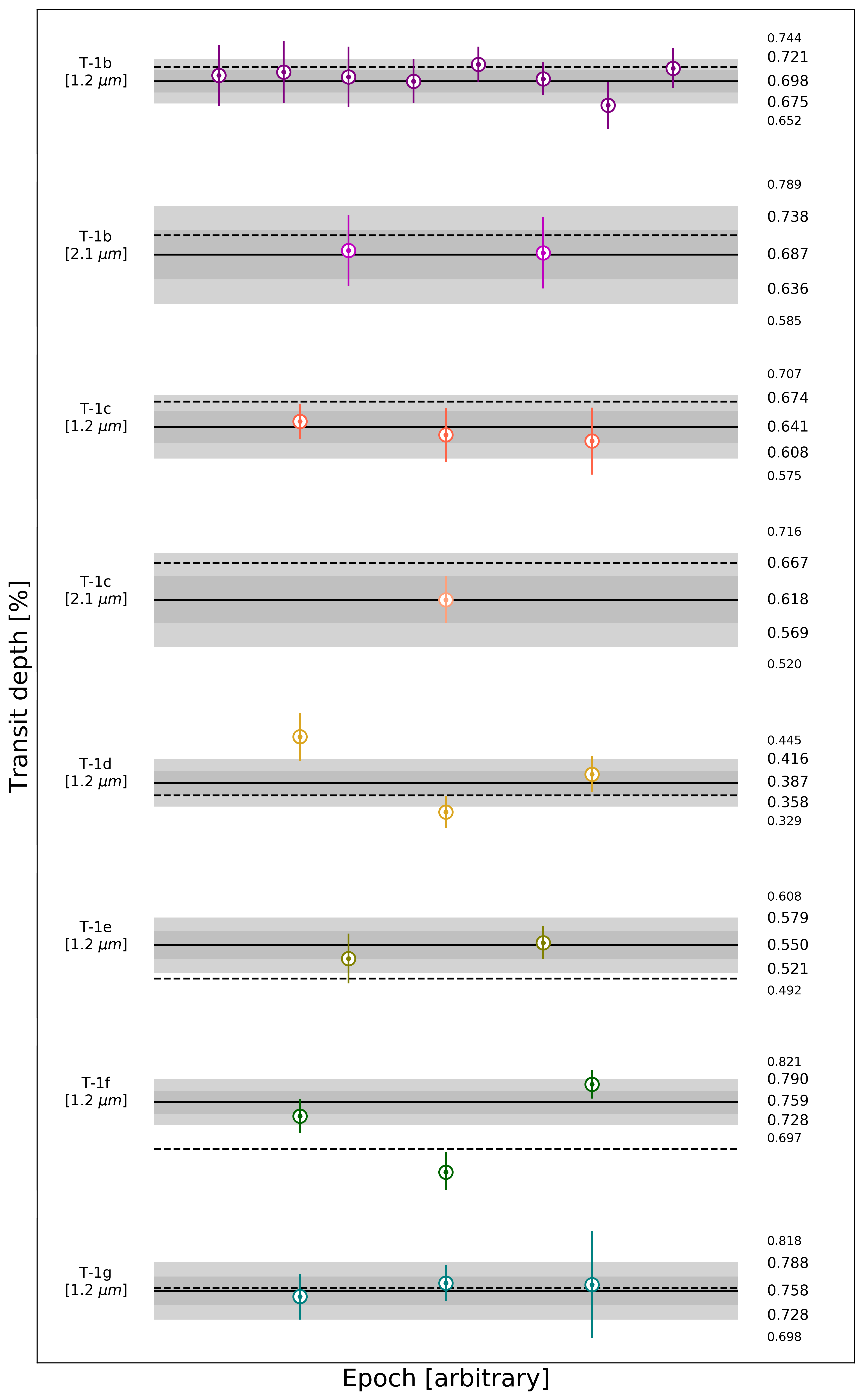}
    \caption{Transit depths measured individually for each transit. Values are displayed chronologically, but not linearly in time. The horizontal black line shows the median value of the transit depth posterior PDF inferred from the global analysis with 1-$\sigma$ and 2-$\sigma$ intervals shown in shades of grey with numerical values on the right. The dotted line represents the same transit depth from the global \textit{Spitzer} data analysis in the thermal-IR 4.5~$\mu$m range \protect\citep{Delrez2018}.}
    \label{df_epoch}
\end{figure}

Transit depths, timings, durations and impact parameters deduced from our analysis are shown in Table~\ref{tab:ind_an} and transit depths are discussed in the \hyperref[RESULTS AND DISCUSSION]{Results and Discussion} section. We plot individual transit depths in Fig.~\ref{df_epoch}. 

\par
\begin{table*}
\small
\caption{Results of the individual analyses: median values of the posterior PDFs and their respective 1-$\sigma$ limits derived for the timing $T_0$ , depth $dF$, duration $W$ and impact parameter $b$ for each transit.}
%\noindent\makebox[\textwidth][c]{
%\begin{minipage}{\textwidth}
\begin{tabular}{cccccccccccc}
\hline
ID & Planet & Epoch & Filter & $T_0$ & $\sigma_{T_0}$ & $dF$ & $\sigma_{dF}$ & $W$ & $\sigma_{W}$ & $b$ & $\sigma_{b}$ \\
\phantom{text} & \phantom{text} & \phantom{text} & \phantom{text} & ($\mathrm{BJD_{TDB}}$ -- 2450000) &  \phantom{text} & (per cent) & \phantom{text} & (min) \\
\hline
%---------------------------------------b---------------------------------------------------------------------
1 & b & 8   & NB2090 & 7334.59940 & 0.00062 & 0.696 & 0.074 & 36.56 & 1.38 & 0.17 & 0.12\\
2 & b & 26  & J & 7361.79958 & 0.00041 & 0.710 & 0.063 & 37.38 & 1.11 & 0.18 & 0.11 \\
3 & b & 28  & J & 7364.82184 &  0.00076 & 0.717 & 0.065 & 35.68 & 1.84 & 0.26 & 0.16\\
4 & b & 183 & J & 7599.00621 &  0.00052 &  0.707 &  0.063 &  36.34 &  1.20 &  0.17 &  0.12\\
5 & b & 185 & J & 7602.02813 & 0.00036 & 0.698 & 0.046 & 37.00 & 1.00 & 0.15 & 0.11\\
6 & b & 236 & J & 7679.08274 & 0.00019 & 0.733 & 0.037 & 36.87 & 0.61 & 0.15 & 0.10 \\
7 & b & 238 & J & 7682.10451 & 0.00022 & 0.703 & 0.034 & 34.46 & 0.72 & 0.28 & 0.12 \\
8 & b & 410 & J & 7941.97621 & 0.00038 & 0.648 & 0.049 & 35.49 & 0.43 & 0.32 & 0.13 \\
9 & b & 507 & NB2090 & 8088.53133 & 0.00056 & 0.691 & 0.074 & 35.67 & 1.28 & 0.23 & 0.14 \\
10 & b & 651 & J & 8306.10046 & 0.00025 & 0.725 & 0.042 & 36.02 & 0.38 & 0.20 & 0.12\\
\\
%----------------c------------------------------------
11 & c & 33  & J & 7362.72643 & 0.00034 & 0.653 & 0.037 & 41.69 & 1.00 & 0.23 & 0.13\\
12 & c & 126 & J & 7587.95740 & 0.00050 & 0.625 & 0.056 & 41.94 & 0.98 & 0.19 & 0.12\\
13 & c & 131  & J & 7600.06699 & 0.00047 & 0.612 & 0.070 & 42.20 & 1.24 & 0.17 & 0.11\\
14 & c & 329 & NB2090 & 8079.58054 & 0.00045 & 0.606 & 0.106 & 42.28 & 1.16 & 0.18 & 0.11\\
\\
%----------------d------------------------------------
15 & d &77  & J & 7981.98734 & 0.00085 & 0.499 & 0.058 & 48.99 & 0.42 & 0.22 & 0.12\\
16 & d &78  & J & 7986.03382 & 0.00050 & 0.316 & 0.039 & 47.88 & 1.33 & 0.23 & 0.13\\
17 & d &95  & J & 8054.87480 & 0.00048 & 0.408 & 0.044 & 50.18 & 1.28 & 0.15 & 0.10\\
\\
%----------------e------------------------------------
18 & e &57  & J & 8008.03130 & 0.00062 & 0.522 & 0.052 & 56.03 & 0.42 & 0.20 & 0.12\\
19 & e &107 & J & 8313.02465 & 0.00035 & 0.555 & 0.034 & 56.80 & 1.04 & 0.18 & 0.11\\
\\
%----------------f------------------------------------
20 & f &32  & J & 7966.01321 & 0.00042 & 0.722 & 0.046 & 61.76 & 1.29 & 0.41 & 0.10\\
21 & f &37  & J & 8012.04130 & 0.00081 & 0.572 & 0.053 & 62.08 & 0.42 & 0.31 & 0.12\\
22 & f &41  & J & 8048.86235 & 0.00026 & 0.807 & 0.038 & 62.04 & 0.39 & 0.35 & 0.11 \\ 
\\
%----------------g------------------------------------
23 & g &27  & J & 7998.88349 & 0.00056 & 0.746 & 0.048 & 69.96 & 1.51 & 0.32 & 0.13\\
24 & g &35  & J & 8097.72511 & 0.00031 & 0.774 & 0.037 & 70.57 & 1.10 & 0.29 & 0.12\\
25 & g &50  & J & 8283.05268 & 0.00093 & 0.771 & 0.111 & 69.14 & 0.43 & 0.37 & 0.12\\
\hline
\end{tabular}
%\end{minipage}}
\label{tab:ind_an}
\end{table*}
\par

By using the approach of combining PDFs of the light curves selected basing on the RMS$\,\times\,\beta_r^2$ metric for every transit, our errors on transit depths tend to increase compared to the analysis of a single 'best' light curve. But this approach gave robust estimates of the transit depths, which is critical in the assessment of their temporal variability and associated possible stellar contamination. 

\subsection{Global analysis}\label{glob_an}

For each planet, we ran a global MCMC analysis of all the transits to check the consistency of the transits depths from the individual MCMC analysis. We used only one light curve from the set of all light curves selected for each transit observation. Said light curve's median value of the posterior PDF for the transit depth was the closest to the median value obtained from the combination of all the PDFs deduced from the set of best light curves for a particular transit. These light curves are presented in \hyperref[A1]{Appendix~A1} along with corresponding RMS versus bin size plots made with the $\mathrm{MC^3}$ software \citep{2017AJ....153....3C} to assess the amount of correlated (red) noise in our time-series. The global analysis was performed for each planet in each filter -- TRAPPIST-1\,b-g in J~band and TRAPPIST-1\,b in NB2090 band. We note that there is only one transit of planet c in NB2090 band.

\begin{figure}
    \centering
    \includegraphics[width=\columnwidth]{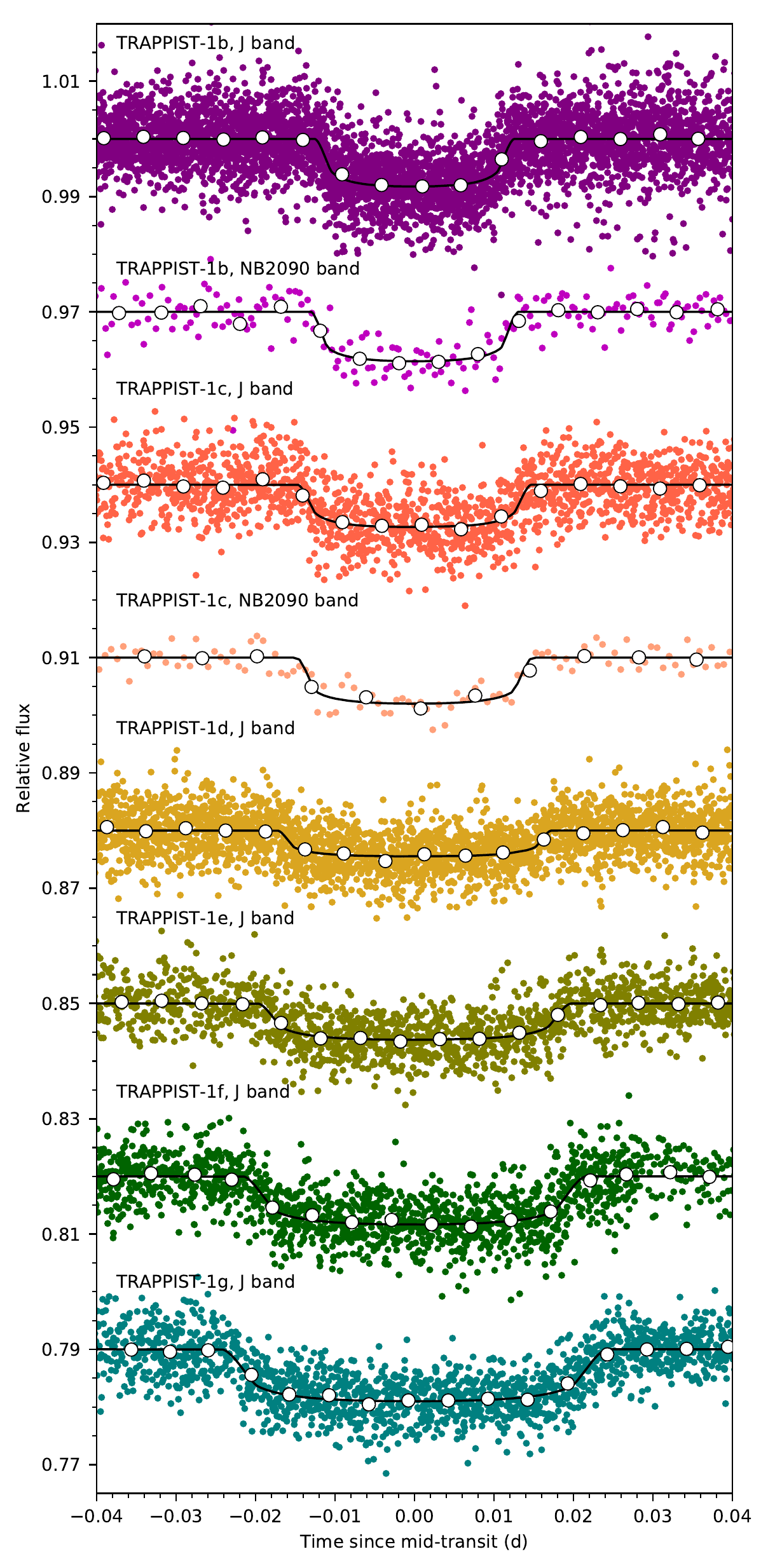}
    \caption{Period-folded transits of TRAPPIST-1\,b-g planets multiplied by the baseline polynomials, and corrected for TTVs. Individual measurements are presented in colored circles and white circles are 7~min binned values. The solid black line represents the best-fit model.}
    \label{fig:phase-folded}
\end{figure}

Jump parameters were the same as for the individual analyses (see Subsection~\ref{ind_an}), but with an addition of a Transit Timing Variation (TTV) for each transit. Period $P$ and initial transit epoch $T_0$ were fixed to the values from \cite{Delrez2018}. Thus there were seven common parameters for all transits (stellar parameters including LD coefficients + transit impact parameter $b$) and two individual parameters for each transit ($dF$ and TTV). We used the same photometric baselines that were applied in the individual analyses. First, we ran MCMC with $10\,000$ steps to estimate the correction factors, CF, and then two chains of $100\,000$ steps with 20\% burn-in phase were executed to derive the transit parameters. Their convergence was checked as well with the test of Gelman \& Rubin. Deduced transit depths are presented in Table~\ref{tab:global_depths} and detrended phase-folded light curves for each planet and bandpass are presented in Fig.~\ref{fig:phase-folded}.

\section{RESULTS AND DISCUSSION}\label{RESULTS AND DISCUSSION}

\subsection{Transit depths variability}

To assess how transit depths change from epoch to epoch, for each planet we compared the depths inferred from the individual analyses with the one measured from the global analysis of all transits (see Table~\ref{tab:ind_an} and Table~\ref{tab:global_depths}, respectively). We conclude that the individual transit depths of TRAPPIST-1\,b, c, e, and g are consistent with the values deduced from the global analysis at better than 1-$\sigma$, and that the maximum standard deviation of the measured individual transit depth does not exceed 260 ppm (the case of TRAPPIST-1\,b, where the depth inferred from light curve \#8 could deviate the most because of a data gap during the egress and a short OOT baseline which prevents us from properly modelling the systematic effects during the observing run). We note that our initial standard deviation of the measured individual transit depth of planet~b, which was obtained using a classical approach to differential photometry by minimization of the OOT RMS, was 1000 ppm. However, even with a rigorous approach, transit depths of TRAPPIST-1\,d and f show temporal variations from epoch to epoch, with peak-to-peak values of 1800 ppm, and 2400~ppm, respectively, and standard deviations are larger than the mean errors.

We note that there are only three transits for planets d and f, the two that appear to have noticeable temporal variations (\cite{2018AJ....156..218D} also reports temporal variations of planet d basing on 10 transits from \textit{K2}). In the case of planet d, all transit observations were done in photometric conditions, but the first transit (light curve \#15) has larger error bars than the others, most likely because of a data gap during an egress. Another transit of TRAPPIST-1\,d suffered from a short data gap in the bottom part (light curve \#16). In each case, after resuming the observations, the telescope pointing held, holding all the stars within 1 pixel from their initial positions, but gaps during the ingress/egress could affect transit shapes, which could affect our deduced transit depths. A similar situation can be seen during a transit of planet~b (above-mentioned light curve \#8), where a data gap is also present during an egress. However in the case of a transit of planet g (light curve \#25) data gap during ingress and short OOT baseline are also present, but the deduced transit depth is consistent with previous measurements (light curves \#23 and \#24). In the case of planet f, the first transit was observed in thin cirrus conditions (light curve \#20), the second transit shows a possible spot-crossing event (light curve \#21) and the third transit suffered from thin cirrus before the ingress (light curve \#22). All transits of TRAPPIST-1\,d and f were observed with \textit{UKIRT} in J band with the same instrumental setup. As precipitable water vapour (PWV) has drastic influence on the opacity of Earth's atmosphere in the near-IR and thus impacts near-IR photometry, we checked each observing night for the amount of PWV\footnote{\url{http://www.eao.hawaii.edu/weather/watervapor/mk/archive/}} and we could not find any correlations between deduced transits depths and the amount of PWV. Besides possible effect of thin cirrus clouds, we could not attribute transit variations in depth to other external factors, such as the position of the star on the chip, mean FWHM, time, airmass or sky background values or any combination of these. Thus we conclude that for now our apparent high scatter of transit depths for planets d and f likely originate from the effects of the data gaps, thin cirrus or the spot-crossing event.

%Simult transits - no overlap with the 1st K2 mission and with Delrez2018. just overlap with SSO and VLT.

\begin{figure}
    \centering
    \includegraphics[width=\columnwidth]{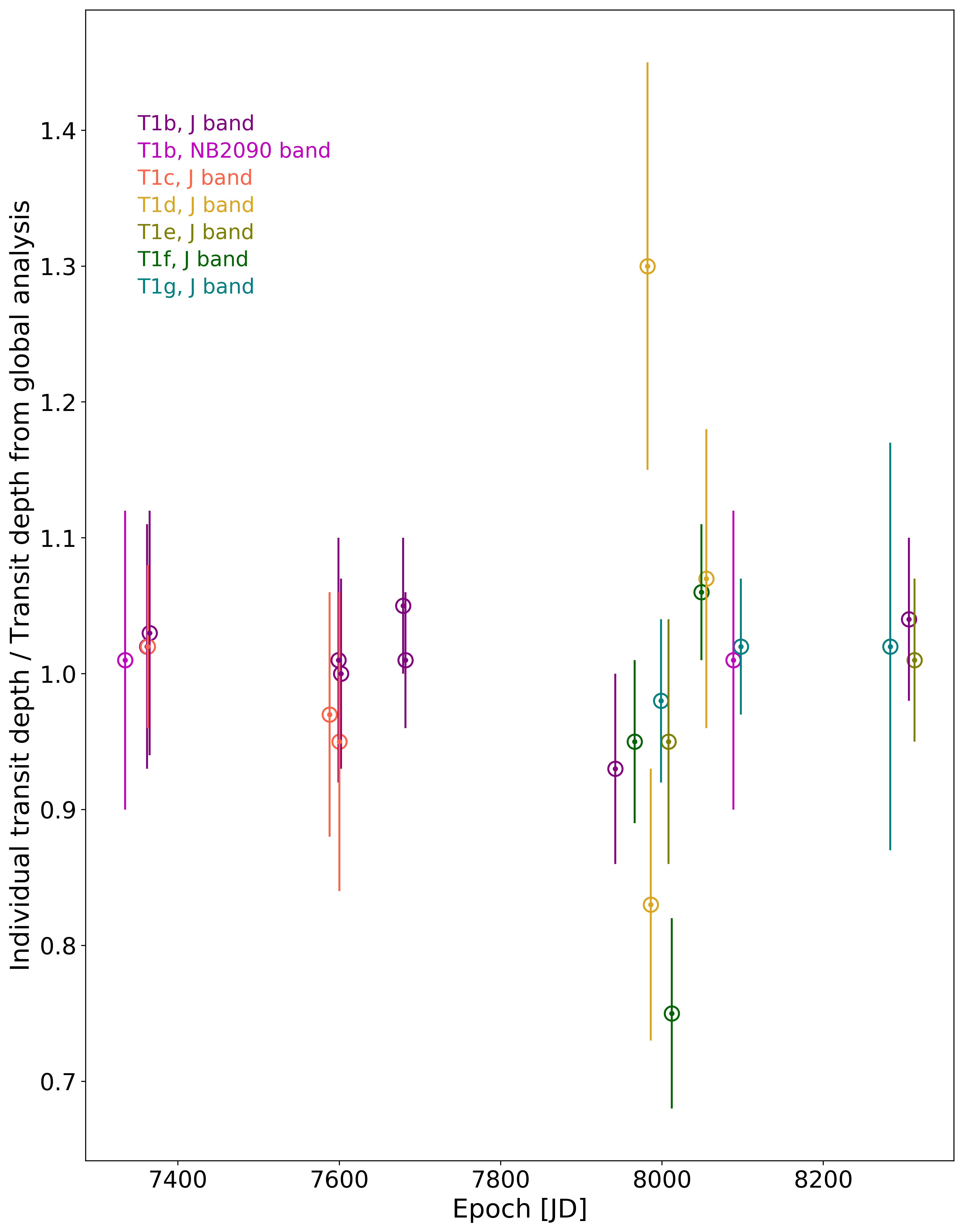}
    \caption{Ratio of the individual transit depth to the average depth from the global MCMC analysis as a function of time (JD).}
    \label{fig:df_jd}
\end{figure}

Interestingly, all transits of TRAPPIST-1\,d and f in the near-IR were obtained in one observing season during Summer and Autumn of 2017. Some of the transits of TRAPPIST-1\,b, c, e, and g were also obtained in the same season, but their values are consistent with observations made in other seasons. In Fig.~\ref{fig:df_jd} we display the ratio of an individual transit depth to the average transit depth inferred from the global analysis for each planet as a function of Julian Date. If the scatter noticed for planets d and f originated from a maximum of the star's magnetic cycle in 2017, we would  expect the transits of the other planets to have been affected too. The fact that it  was not the case argues against this hypothesis.    

All our observations were centered only on transit windows with the longest observing window lasting 5~h which makes it hard to properly sample the rotational period of the star, reported to be $\sim$3.3~d basing on the \textit{K2} data \citep{Luger2017}. However, this period could be a characteristic timescale of active regions \citep{2018ApJ...857...39M}. In any case, more transit observations of TRAPPIST-1 planets, especially of planets d and f, and frequent photometric monitoring of the host star are needed to confirm astrophysical origin and understand the real cause of transit depth variations of TRAPPIST-1\,d and f. 

\begin{table}
    \centering
    \caption{Transit depths from the global analyses: median values of the posterior PDFs and their respective 1-$\sigma$ limits.}
    \begin{tabular}{c|c|c}
    \hline
         Planet & $dF_\mathrm{J}$ & $dF_{\mathrm{NB2090}}$\\
         \phantom{text} & (per cent) & (per cent)\\
    \hline
         TRAPPIST-1\,b & $0.700\pm0.023$ & $0.688\pm0.05$\\
         TRAPPIST-1\,c & $0.641\pm0.034$ & $0.618\pm0.04$\\
         TRAPPIST-1\,d & $0.387\pm0.029$ & -\\
         TRAPPIST-1\,e & $0.550\pm0.029$ & -\\
         TRAPPIST-1\,f & $0.759\pm0.031$ & -\\
         TRAPPIST-1\,g & $0.758\pm0.030$ & -\\
    \hline
    \end{tabular}
    \label{tab:global_depths}
\end{table}

\begin{table}
    \centering
    \caption{Standard deviations and mean errors of the transit depths from the individual analyses.}
    \begin{tabular}{c|c|c|c}
        \hline
        Planet & Filter & $\sigma$ & Mean error\\
        \phantom{text} & \phantom{text} & (per cent) & (per cent)\\ 
        \hline
         TRAPPIST-1\,b & J & 0.026 & 0.050\\
         TRAPPIST-1\,b & NB2090 & 0.004 & 0.074\\
         TRAPPIST-1\,c & J & 0.021 & 0.054\\
         TRAPPIST-1\,d & J & 0.091 & 0.047\\
         TRAPPIST-1\,e & J & 0.023 & 0.043\\
         TRAPPIST-1\,f & J & 0.119 & 0.046\\
         TRAPPIST-1\,g & J & 0.015 & 0.066\\
        \hline
    \end{tabular}
    \label{tab:depth_stdv}
\end{table}

\subsection{Updated transmission spectra and stellar contamination}

\begin{figure}
    \centering
    \includegraphics[width=\columnwidth]{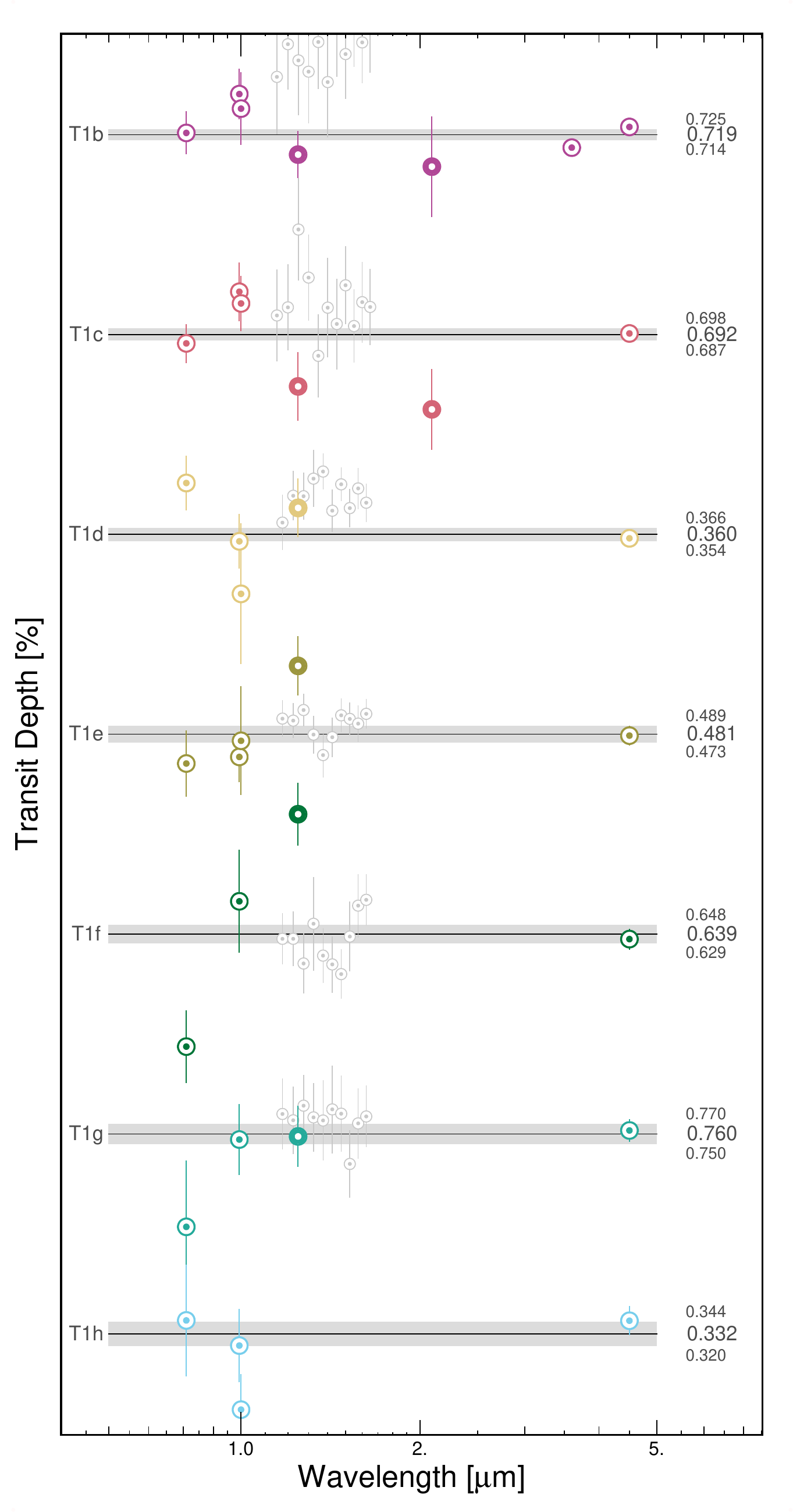}
    \caption{Transmission spectra of TRAPPIST-1\,b-h planets. The solid black line is the weighted mean of all measurements excluding \textit{HST} presented in \protect\cite{2018AJ....156..218D} with corresponding 1-$\sigma$ confidence intervals in shades of gray and numerical values on the right. \textit{HST} measurements are presented as grey points. Measurements in the near-IR (1.2~$\micron$ and 2.1~$\micron$) from this study are presented as colored circles with white center. Each point is a median value of the posterior PDF and its respective 1-$\sigma$ from the global analysis at the effective wavelength of corresponding instrument.}
    \label{fig:transm}
\end{figure}

We updated the transmission spectra of TRAPPIST-1\,b-g planets presented in \cite{2018AJ....156..218D} by adding transit depths deduced from our global analysis in J and NB2090 bands. Our data set only includes two NB2090 band transits of TRAPPIST-1\,b and one of TRAPPIST-1\,c which is a cleaner, narrower bandpass (see Subsection~\ref{subsec:vlt}). Additional transit depth measurements are especially needed with this bandpass to improve the precision. The updated transmission spectra is presented in Fig.~\ref{fig:transm}. 

We note that transit depths of planet b in the near-IR are consistent with previously published non-\textit{HST} transit depth measurements. They lie in most cases within $\sim$ 1-$\sigma$ of the \textit{HST} results, but are consistently shallower in depth. \textit{HST} measurements in their turn could be affected by the telescope orbit-dependent systematic effects, which could result in much deeper transit depths (see \cite{deWit2016}, Subsection 3.2 in \cite{2018AJ....156..218D} and \cite{Zhang2018} for details). For TRAPPIST-1\,c our deduced transit depths also disagree somewhat with \textit{HST} measurements and show, like TRAPPIST-1\,b, shallower transits comparing to \textit{Spitzer} 4.5~$\mu$m depths, but lie within the 3-$\sigma$ interval. Transit depths of TRAPPIST-1\,d are in agreement with \textit{HST} data, and show excellent agreement for g. For planets e and f our deduced depths in 1.2~~$\mu$m are deeper than all published depths in other spectral ranges. But these findings are based on just two transits of TRAPPIST-1\,e, and transits of TRAPPIST-1\,f show considerable temporal variations. 

Though there is a certain difficulty in obtaining absolute near-IR transit depths, it may be mitigated by using different instruments with partially overlapping wavelength bands. This mitigation procedure requires spectra taken with different instruments requires simultaneous multi-instruments observations in order to avoid compensating for time- and wavelength-dependent effects that may contaminate a planetary transmission spectrum, such as the epoch-dependent brightness distribution of the host star.

\begin{figure}
    \centering
    \includegraphics[width=\columnwidth]{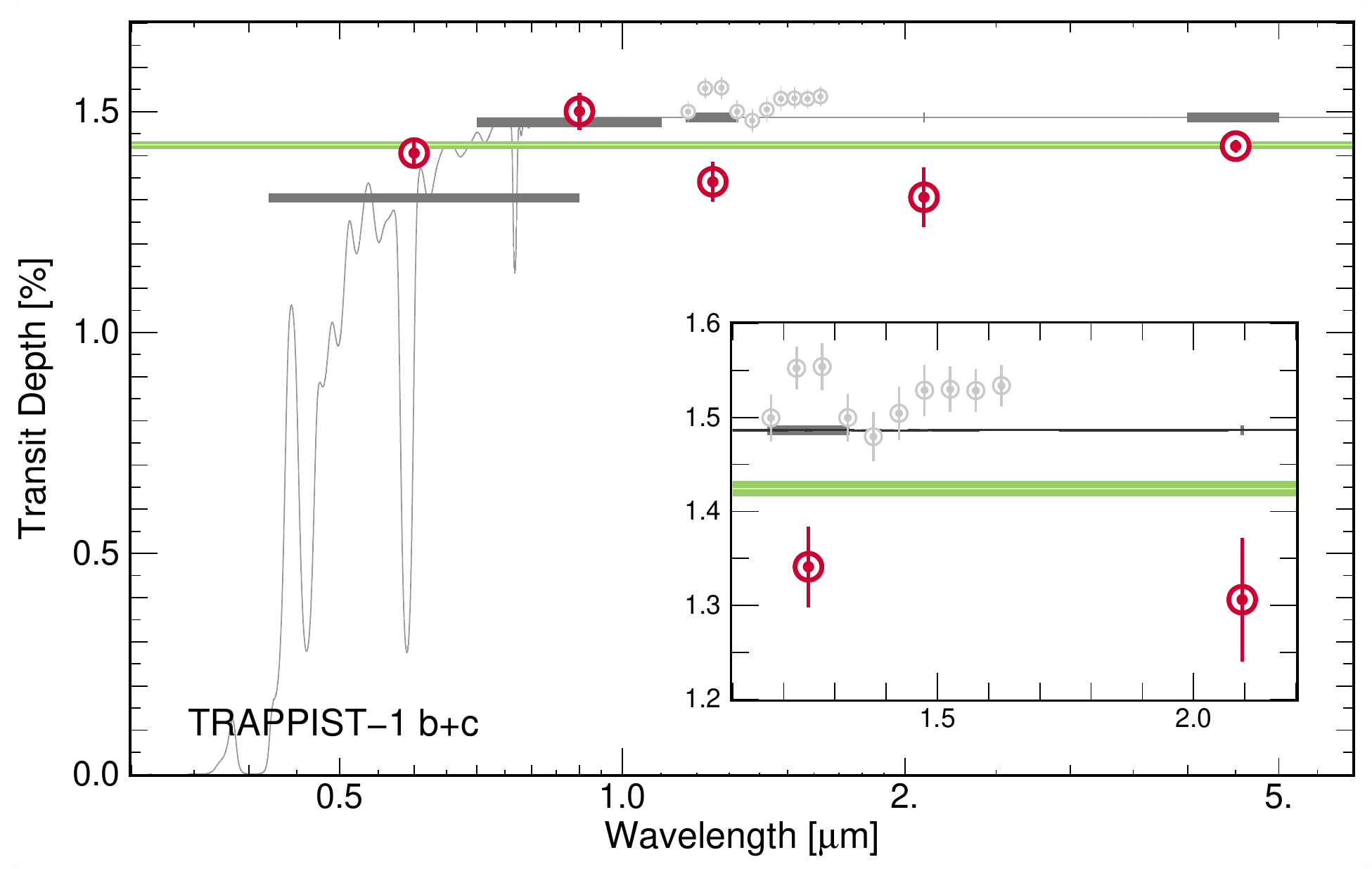}
    \caption{Stellar contamination spectra proposed by \protect\cite{2018ApJ...857...39M} originating from possible existence of bright spots of temperature 4500~K (gray continuous line) with plotted observed depth variations displayed as red points (depths from \textit{HST} are presented as grey points). The gray horizontal bars are the band-integrated value for stellar contamination spectra where the integrals are weighted uniformly in wavelength.}
    \label{fig:contam}
\end{figure}

We also added our depth measurements for planets b and c to the stellar contamination spectra proposed by \cite{2018ApJ...857...39M}, which originate from the possible existence of bright spots of temperature 4500~K. We find that the prediction implies a flat contamination spectrum for wavelengths redder than 0.7~$\mu$m and that this model disagrees with what we see in near-IR (see Fig.~\ref{fig:contam}). Since the real distribution of the spots and their temperatures are unknown, the TRAPPIST-1 system would benefit from additional monitoring in the near-IR to put better constraints on contamination models. In addition, current contamination models only account for spots or faculaes or their combinations, but the atmosphere of TRAPPIST-1 could contain dust. \cite{2019MNRAS.484L..38M} using linear polarization photometry in the near-IR J band finds hints that the atmosphere of TRAPPIST-1 is quite dusty. Likely, this should also be included for proper accounting of stellar contamination.

%\begin{figure}
%    \centering
%    \includegraphics[width=\columnwidth]{cont_1.png}
%    \caption{}
%    \label{fig:df_jd}
%\end{figure}

\section{CONCLUSIONS}\label{CONCL}

We presented here an extensive photometric data set of 25 transits of TRAPPIST-1 observed in the near-IR with \textit{UKIRT} and \textit{AAT} in J band (1.2~$\mu$m) and with the \textit{VLT} in the NB2090 band (2.1~~$\mu$m) from 2015 to 2018. We deduced individual transit depths for each transit taking into account inherent to ground-based near-IR observations correlations of transit parameters with the selected sets of comparison stars and photometric aperture sizes to obtain results as robust and uniform as possible.

We reach a photometric precision of $\sim0.003$ (RMS of the residuals), and we detect no significant temporal variations of transit depths of TRAPPIST-1\,b, c, e, and g, while transit depths of planets d and f show hints of variability with peak-to-peak values of 1800 and 2400~ppm, respectively. Besides thin cirrus observing conditions, we could not link these depth variations to any external parameters, including the amount of precipitable water vapour, the position of the star on the detector, the mean FWHM, etc. We deduce that they likely originate from a few abnormal transits that were affected by thin cirrus, data gaps or, for one, by a possible spot-crossing event. Considering the small number of observed transits for TRAPPIST-1\,d and f, the system would benefit from more transit observations and from frequent monitoring in the near-IR to probe photometric activity of the host star, to better characterize its photospheric homogeneity and to understand if hints of temporal variability are of astrophysical origin. 

We did not detect any flare nor transit that could be attributed to other undetected planets. One possible clear spot crossing event of TRAPPIST-1\,f is presented in light curve \#21 with an amplitude of 200~ppm. We also computed transit timings which will be helpful for further mass and densities updates of the planets via TTV studies. Our depth measurements for planets b and c disagree with the stellar contamination spectra proposed by \cite{2018ApJ...857...39M}, which originate from the possible existence of bright spots of temperature 4500~K. Finally, updated transmission spectra are presented for the six inner planets of the system. We conclude that spectra of TRAPPIST-1\,b and g are globally flat, but some structures are seen for planets c, d, e and f what could be epoch-dependent and coming from the stellar origin.

%\textbf{The near-IR spectral range encompasses the peak of the spectral energy distribution of TRAPPIST-1 and our observations could bring strong constraints on the photospheric properties of the host star and on the impact of stellar contamination to be expected for upcoming \textit{JWST} observations of the planets.}

\section*{Acknowledgements}
\label{Ack}

TRAPPIST-1 observations in NB2090 band are based on observations collected at the European Southern Observatory under ESO programmes 296.C-5010(A) and 598.C-0347(A). Observations with  UKIRT were taken through non-survey projects U/15B/NA05A, U/16A/NA04, U/16B/NA03, U/17A/NA01, U/17B/NA01, U/18A/NAV05, and U/18B/NAV03. Observations with the AAT are based on programmes A/2016B/109 and A/2017B/104.

Computational resources have been provided by the Consortium des {\'E}quipements de Calcul Intensif (C{\'E}CI), funded by the Fonds de la Recherche Scientifique de Belgique (F.R.S.-FNRS) under Grant No. 2.5020.11 and by the Walloon Region.

PyRAF is a product of the Space Telescope Science Institute, which is operated by AURA for NASA.

Authors would like to thank Mike Road and Mike Irwin from CASU for their help with access to the UKIRT data at WSA; Staff Scientists Drs. Tom Kerr and Watson Varricatt, Telescope Operators Sam Benigni, Tim Carroll, Eric Moore, and Michael Pohlen for scheduling and performing observations with the \textit{UKIRT}; NASA, University of Arizona, University of Hawai'i, Lockheed Martin, and the US Naval Observatory for support and granting time for the above noted observing programs; Duncan Wright and Tayyaba Zafar for performing observations with the \textit{AAT}; Evgeny Ivanov for his advises on (C{\'E}CI) computations.
 
The research leading to these results have received funding from the ARC grant for Concerted Research Actions, financed by the Wallonia-Brussels Federation, and by the NASA K2 Cycle 4 program.

V.V.G. is a F.R.S.-FNRS Research Associate, M.G. and E.J. are F.R.S.-FNRS Senior Research Associates. 

Authors would like to thank the anonymous reviewer for the constructive comments, which helped us to improve the quality of the paper.

%%%%%%%%%%%%%%%%%%%%%%%%%%%%%%%%%%%%%%%%%%%%%%%%%%



\bibliographystyle{mnras}
\bibliography{Burdanov}

\begin{thebibliography}{}
\makeatletter
\relax
\def\mn@urlcharsother{\let\do\@makeother \do\$\do\&\do\#\do\^\do\_\do\%\do\~}
\def\mn@doi{\begingroup\mn@urlcharsother \@ifnextchar [ {\mn@doi@}
  {\mn@doi@[]}}
\def\mn@doi@[#1]#2{\def\@tempa{#1}\ifx\@tempa\@empty \href
  {http://dx.doi.org/#2} {doi:#2}\else \href {http://dx.doi.org/#2} {#1}\fi
  \endgroup}
\def\mn@eprint#1#2{\mn@eprint@#1:#2::\@nil}
\def\mn@eprint@arXiv#1{\href {http://arxiv.org/abs/#1} {{\tt arXiv:#1}}}
\def\mn@eprint@dblp#1{\href {http://dblp.uni-trier.de/rec/bibtex/#1.xml}
  {dblp:#1}}
\def\mn@eprint@#1:#2:#3:#4\@nil{\def\@tempa {#1}\def\@tempb {#2}\def\@tempc
  {#3}\ifx \@tempc \@empty \let \@tempc \@tempb \let \@tempb \@tempa \fi \ifx
  \@tempb \@empty \def\@tempb {arXiv}\fi \@ifundefined
  {mn@eprint@\@tempb}{\@tempb:\@tempc}{\expandafter \expandafter \csname
  mn@eprint@\@tempb\endcsname \expandafter{\@tempc}}}

\bibitem[\protect\citeauthoryear{{Apai} et~al.,}{{Apai} et~al.}{2018}]{Apai}
{Apai} D.,  et~al., 2018, preprint, \href
  {http://adsabs.harvard.edu/abs/2018arXiv180308708A} {} (\mn@eprint {arXiv}
  {1803.08708})

\bibitem[\protect\citeauthoryear{Barstow \& Irwin}{Barstow \&
  Irwin}{2016}]{Barstow2016}
Barstow J.~K.,  Irwin P. G.~J.,  2016, \mn@doi [Monthly Notices of the Royal
  Astronomical Society: Letters] {10.1093/mnrasl/slw109}, 461, L92

\bibitem[\protect\citeauthoryear{{Casali} et~al.,}{{Casali}
  et~al.}{2007}]{2007A&A...467..777C}
{Casali} M.,  et~al., 2007, \mn@doi [\aap] {10.1051/0004-6361:20066514}, \href
  {http://adsabs.harvard.edu/abs/2007A%26A...467..777C} {467, 777}

\bibitem[\protect\citeauthoryear{{Claret}, {Hauschildt}  \& {Witte}}{{Claret}
  et~al.}{2012}]{2012A&A...546A..14C}
{Claret} A.,  {Hauschildt} P.~H.,   {Witte} S.,  2012, \mn@doi [\aap]
  {10.1051/0004-6361/201219849}, \href
  {http://adsabs.harvard.edu/abs/2012A%26A...546A..14C} {546, A14}

\bibitem[\protect\citeauthoryear{{Clark}, {Anderson}, {Madhusudhan}, {Hellier},
  {Smith}  \& {Collier Cameron}}{{Clark} et~al.}{2018}]{2018A&A...615A..86C}
{Clark} B.~J.~M.,  {Anderson} D.~R.,  {Madhusudhan} N.,  {Hellier} C.,  {Smith}
  A.~M.~S.,   {Collier Cameron} A.,  2018, \mn@doi [\aap]
  {10.1051/0004-6361/201527071}, \href
  {http://adsabs.harvard.edu/abs/2018A%26A...615A..86C} {615, A86}

\bibitem[\protect\citeauthoryear{{Croll} et~al.,}{{Croll}
  et~al.}{2015}]{2015ApJ...802...28C}
{Croll} B.,  et~al., 2015, \mn@doi [\apj] {10.1088/0004-637X/802/1/28}, \href
  {http://adsabs.harvard.edu/abs/2015ApJ...802...28C} {802, 28}

\bibitem[\protect\citeauthoryear{{Cubillos}, {Harrington}, {Loredo}, {Lust},
  {Blecic}  \& {Stemm}}{{Cubillos} et~al.}{2017}]{2017AJ....153....3C}
{Cubillos} P.,  {Harrington} J.,  {Loredo} T.~J.,  {Lust} N.~B.,  {Blecic} J.,
   {Stemm} M.,  2017, \mn@doi [\aj] {10.3847/1538-3881/153/1/3}, \href
  {https://ui.adsabs.harvard.edu/\#abs/2017AJ....153....3C} {153, 3}

\bibitem[\protect\citeauthoryear{{Delrez} et~al.,}{{Delrez}
  et~al.}{2018}]{Delrez2018}
{Delrez} L.,  et~al., 2018, \mn@doi [\mnras] {10.1093/mnras/sty051}, \href
  {http://adsabs.harvard.edu/abs/2018MNRAS.475.3577D} {475, 3577}

\bibitem[\protect\citeauthoryear{{Ducrot} et~al.,}{{Ducrot}
  et~al.}{2018}]{2018AJ....156..218D}
{Ducrot} E.,  et~al., 2018, \mn@doi [\aj] {10.3847/1538-3881/aade94}, \href
  {http://adsabs.harvard.edu/abs/2018AJ....156..218D} {156, 218}

\bibitem[\protect\citeauthoryear{{Gelman} \& {Rubin}}{{Gelman} \&
  {Rubin}}{1992}]{1992StaSc...7..457G}
{Gelman} A.,  {Rubin} D.~B.,  1992, \mn@doi [Statistical Science]
  {10.1214/ss/1177011136}, \href
  {http://adsabs.harvard.edu/abs/1992StaSc...7..457G} {7, 457}

\bibitem[\protect\citeauthoryear{{Gillon} et~al.,}{{Gillon}
  et~al.}{2012}]{2012A&A...542A...4G}
{Gillon} M.,  et~al., 2012, \mn@doi [\aap] {10.1051/0004-6361/201218817}, \href
  {http://adsabs.harvard.edu/abs/2012A%26A...542A...4G} {542, A4}

\bibitem[\protect\citeauthoryear{{Gillon} et~al.,}{{Gillon}
  et~al.}{2014}]{2014A&A...563A..21G}
{Gillon} M.,  et~al., 2014, \mn@doi [\aap] {10.1051/0004-6361/201322362}, \href
  {http://adsabs.harvard.edu/abs/2014A%26A...563A..21G} {563, A21}

\bibitem[\protect\citeauthoryear{{Gillon} et~al.,}{{Gillon}
  et~al.}{2016}]{Gillon2016}
{Gillon} M.,  et~al., 2016, \mn@doi [Nature] {10.1038/nature17448}, \href
  {http://adsabs.harvard.edu/abs/2016Natur.533..221G} {533, 221}

\bibitem[\protect\citeauthoryear{Gillon et~al.,}{Gillon
  et~al.}{2017}]{Gillon2017}
Gillon M.,  et~al., 2017, \mn@doi [Nature] {10.1038/nature21360}, 542, 456

\bibitem[\protect\citeauthoryear{{Gizis}, {Monet}, {Reid}, {Kirkpatrick},
  {Liebert}  \& {Williams}}{{Gizis} et~al.}{2000}]{2000AJ....120.1085G}
{Gizis} J.~E.,  {Monet} D.~G.,  {Reid} I.~N.,  {Kirkpatrick} J.~D.,  {Liebert}
  J.,   {Williams} R.~J.,  2000, \mn@doi [\aj] {10.1086/301456}, \href
  {http://adsabs.harvard.edu/abs/2000AJ....120.1085G} {120, 1085}

\bibitem[\protect\citeauthoryear{{Grimm} et~al.,}{{Grimm}
  et~al.}{2018}]{Grimm2018}
{Grimm} S.~L.,  et~al., 2018, \mn@doi [\aap] {10.1051/0004-6361/201732233},
  \href {http://adsabs.harvard.edu/abs/2018A%26A...613A..68G} {613, A68}

\bibitem[\protect\citeauthoryear{{Howell}}{{Howell}}{2006}]{2006hca..book.....H}
{Howell} S.~B.,  2006, {Handbook of CCD Astronomy}

\bibitem[\protect\citeauthoryear{{Lendl}, {Gillon}, {Queloz}, {Alonso},
  {Fumel}, {Jehin}  \& {Naef}}{{Lendl} et~al.}{2013}]{2013A&A...552A...2L}
{Lendl} M.,  {Gillon} M.,  {Queloz} D.,  {Alonso} R.,  {Fumel} A.,  {Jehin} E.,
    {Naef} D.,  2013, \mn@doi [\aap] {10.1051/0004-6361/201220924}, \href
  {https://ui.adsabs.harvard.edu/\#abs/2013A&A...552A...2L} {552, A2}

\bibitem[\protect\citeauthoryear{{Lincowski}, {Meadows}, {Crisp}, {Robinson},
  {Luger}, {Lustig-Yaeger}  \& {Arney}}{{Lincowski}
  et~al.}{2018}]{2018ApJ...867...76L}
{Lincowski} A.~P.,  {Meadows} V.~S.,  {Crisp} D.,  {Robinson} T.~D.,  {Luger}
  R.,  {Lustig-Yaeger} J.,   {Arney} G.~N.,  2018, \mn@doi [\apj]
  {10.3847/1538-4357/aae36a}, \href
  {http://adsabs.harvard.edu/abs/2018ApJ...867...76L} {867, 76}

\bibitem[\protect\citeauthoryear{{Luger} et~al.,}{{Luger}
  et~al.}{2017}]{Luger2017}
{Luger} R.,  et~al., 2017, \mn@doi [Nature Astronomy]
  {10.1038/s41550-017-0129}, \href
  {http://adsabs.harvard.edu/abs/2017NatAs...1E.129L} {1, 0129}

\bibitem[\protect\citeauthoryear{{Mandel} \& {Agol}}{{Mandel} \&
  {Agol}}{2002}]{2002ApJ...580L.171M}
{Mandel} K.,  {Agol} E.,  2002, \mn@doi [\apjl] {10.1086/345520}, \href
  {http://adsabs.harvard.edu/abs/2002ApJ...580L.171M} {580, L171}

\bibitem[\protect\citeauthoryear{{Miles-P{\'a}ez}, {Zapatero Osorio},
  {Pall{\'e}}  \& {Metchev}}{{Miles-P{\'a}ez}
  et~al.}{2019}]{2019MNRAS.484L..38M}
{Miles-P{\'a}ez} P.~A.,  {Zapatero Osorio} M.~R.,  {Pall{\'e}} E.,   {Metchev}
  S.~A.,  2019, \mn@doi [\mnras] {10.1093/mnrasl/slz001}, \href
  {http://adsabs.harvard.edu/abs/2019MNRAS.484L..38M} {484, L38}

\bibitem[\protect\citeauthoryear{{Moran}, {H{\"o}rst}, {Batalha}, {Lewis}  \&
  {Wakeford}}{{Moran} et~al.}{2018}]{2018AJ....156..252M}
{Moran} S.~E.,  {H{\"o}rst} S.~M.,  {Batalha} N.~E.,  {Lewis} N.~K.,
  {Wakeford} H.~R.,  2018, \mn@doi [\aj] {10.3847/1538-3881/aae83a}, \href
  {http://adsabs.harvard.edu/abs/2018AJ....156..252M} {156, 252}

\bibitem[\protect\citeauthoryear{{Morley}, {Kreidberg}, {Rustamkulov},
  {Robinson}  \& {Fortney}}{{Morley} et~al.}{2017}]{Morley2017}
{Morley} C.~V.,  {Kreidberg} L.,  {Rustamkulov} Z.,  {Robinson} T.,   {Fortney}
  J.~J.,  2017, \mn@doi [\apj] {10.3847/1538-4357/aa927b}, \href
  {http://adsabs.harvard.edu/abs/2017ApJ...850..121M} {850, 121}

\bibitem[\protect\citeauthoryear{{Morris}, {Agol}, {Davenport}  \&
  {Hawley}}{{Morris} et~al.}{2018a}]{2018ApJ...857...39M}
{Morris} B.~M.,  {Agol} E.,  {Davenport} J.~R.~A.,   {Hawley} S.~L.,  2018a,
  \mn@doi [\apj] {10.3847/1538-4357/aab6a5}, \href
  {http://adsabs.harvard.edu/abs/2018ApJ...857...39M} {857, 39}

\bibitem[\protect\citeauthoryear{{Morris} et~al.,}{{Morris}
  et~al.}{2018b}]{2018ApJ...863L..32M}
{Morris} B.~M.,  et~al., 2018b, \mn@doi [\apjl] {10.3847/2041-8213/aad8aa},
  \href {http://adsabs.harvard.edu/abs/2018ApJ...863L..32M} {863, L32}

\bibitem[\protect\citeauthoryear{{Rackham}, {Apai}  \& {Giampapa}}{{Rackham}
  et~al.}{2018}]{Rackham2018}
{Rackham} B.~V.,  {Apai} D.,   {Giampapa} M.~S.,  2018, \mn@doi [\apj]
  {10.3847/1538-4357/aaa08c}, \href
  {http://adsabs.harvard.edu/abs/2018ApJ...853..122R} {853, 122}

\bibitem[\protect\citeauthoryear{{Rodler} \& {L{\'o}pez-Morales}}{{Rodler} \&
  {L{\'o}pez-Morales}}{2014}]{2014ApJ...781...54R}
{Rodler} F.,  {L{\'o}pez-Morales} M.,  2014, \mn@doi [\apj]
  {10.1088/0004-637X/781/1/54}, \href
  {http://adsabs.harvard.edu/abs/2014ApJ...781...54R} {781, 54}

\bibitem[\protect\citeauthoryear{{Schwarz}}{{Schwarz}}{1978}]{1978AnSta...6..461S}
{Schwarz} G.,  1978, Annals of Statistics, \href
  {http://adsabs.harvard.edu/abs/1978AnSta...6..461S} {6, 461}

\bibitem[\protect\citeauthoryear{{Siebenmorgen}, {Carraro}, {Valenti},
  {Petr-Gotzens}, {Brammer}, {Garcia}  \& {Casali}}{{Siebenmorgen}
  et~al.}{2011}]{2011Msngr.144....9S}
{Siebenmorgen} R.,  {Carraro} G.,  {Valenti} E.,  {Petr-Gotzens} M.,  {Brammer}
  G.,  {Garcia} E.,   {Casali} M.,  2011, The Messenger, \href
  {http://esoads.eso.org/abs/2011Msngr.144....9S} {144, 9}

\bibitem[\protect\citeauthoryear{{Tinney} et~al.,}{{Tinney}
  et~al.}{2004}]{2004SPIE.5492..998T}
{Tinney} C.~G.,  et~al., 2004, in {Moorwood} A.~F.~M.,  {Iye} M.,  eds,
  \procspie Vol. 5492, Ground-based Instrumentation for Astronomy. pp
  998--1009, \mn@doi{10.1117/12.550980}

\bibitem[\protect\citeauthoryear{{Tody}}{{Tody}}{1986}]{Tody1986}
{Tody} D.,  1986, in {Crawford} D.~L.,  ed.,  Proceedings of the Meeting,
  Tucson, AZ, March 4-8, 1986 Vol. 627, Instrumentation in astronomy VI.
  Society of Photo-Optical Instrumentation Engineers (SPIE) Conference Series,
  Bellingham, WA, p.~733

\bibitem[\protect\citeauthoryear{{Van Grootel} et~al.,}{{Van Grootel}
  et~al.}{2018}]{2018ApJ...853...30V}
{Van Grootel} V.,  et~al., 2018, \mn@doi [\apj] {10.3847/1538-4357/aaa023},
  \href {http://adsabs.harvard.edu/abs/2018ApJ...853...30V} {853, 30}

\bibitem[\protect\citeauthoryear{{Wakeford} et~al.,}{{Wakeford}
  et~al.}{2019}]{2019AJ....157...11W}
{Wakeford} H.~R.,  et~al., 2019, \mn@doi [\aj] {10.3847/1538-3881/aaf04d},
  \href {http://adsabs.harvard.edu/abs/2019AJ....157...11W} {157, 11}

\bibitem[\protect\citeauthoryear{{Winn} et~al.,}{{Winn}
  et~al.}{2008}]{2008ApJ...683.1076W}
{Winn} J.~N.,  et~al., 2008, \mn@doi [\apj] {10.1086/589737}, \href
  {http://adsabs.harvard.edu/abs/2008ApJ...683.1076W} {683, 1076}

\bibitem[\protect\citeauthoryear{{Zhang}, {Zhou}, {Rackham}  \& {Apai}}{{Zhang}
  et~al.}{2018}]{Zhang2018}
{Zhang} Z.,  {Zhou} Y.,  {Rackham} B.,   {Apai} D.,  2018, preprint, \href
  {http://adsabs.harvard.edu/abs/2018arXiv180202086Z} {} (\mn@eprint {arXiv}
  {1802.02086})

\bibitem[\protect\citeauthoryear{{de Wit} et~al.,}{{de Wit}
  et~al.}{2016}]{deWit2016}
{de Wit} J.,  et~al., 2016, \mn@doi [\nat] {10.1038/nature18641}, \href
  {http://adsabs.harvard.edu/abs/2016Natur.537...69D} {537, 69}

\bibitem[\protect\citeauthoryear{{de Wit} et~al.,}{{de Wit}
  et~al.}{2018}]{deWit2018}
{de Wit} J.,  et~al., 2018, \mn@doi [Nature Astronomy]
  {10.1038/s41550-017-0374-z}, \href
  {http://adsabs.harvard.edu/abs/2018NatAs...2..214D} {2, 214}

\makeatother
\end{thebibliography}

%%%%%%%%%%%%%%%%%%%%%%%%%%%%%%%%%%%%%%%%%%%%%%%%%%

%%%%%%%%%%%%%%%%% APPENDICES %%%%%%%%%%%%%%%%%%%%%

\appendix

\section{light curves and respective binned residuals RMS versus bin size plots}\label{A1}
\onecolumn

\begin{figure}
    \centering
    \includegraphics[width=0.95\textwidth]{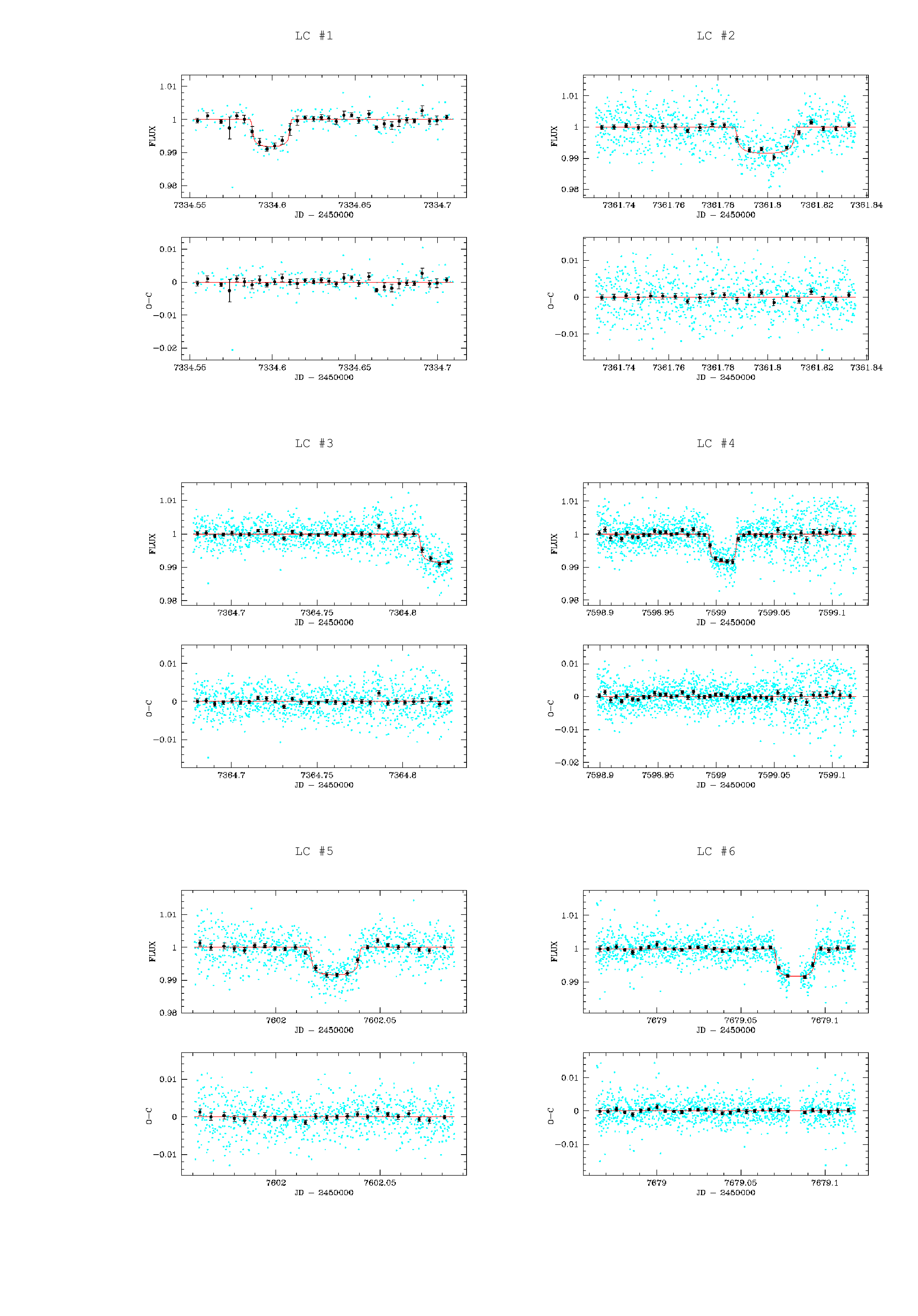}
    \caption{light curves (LC) \#1-6. See Tables \ref{tab:sets},\ref{tab:baselines}, and \ref{tab:ind_an} for the details.}
    \label{fig:plot1}
\end{figure}

\clearpage

\begin{figure}
    \centering
    \includegraphics[width=0.95\textwidth]{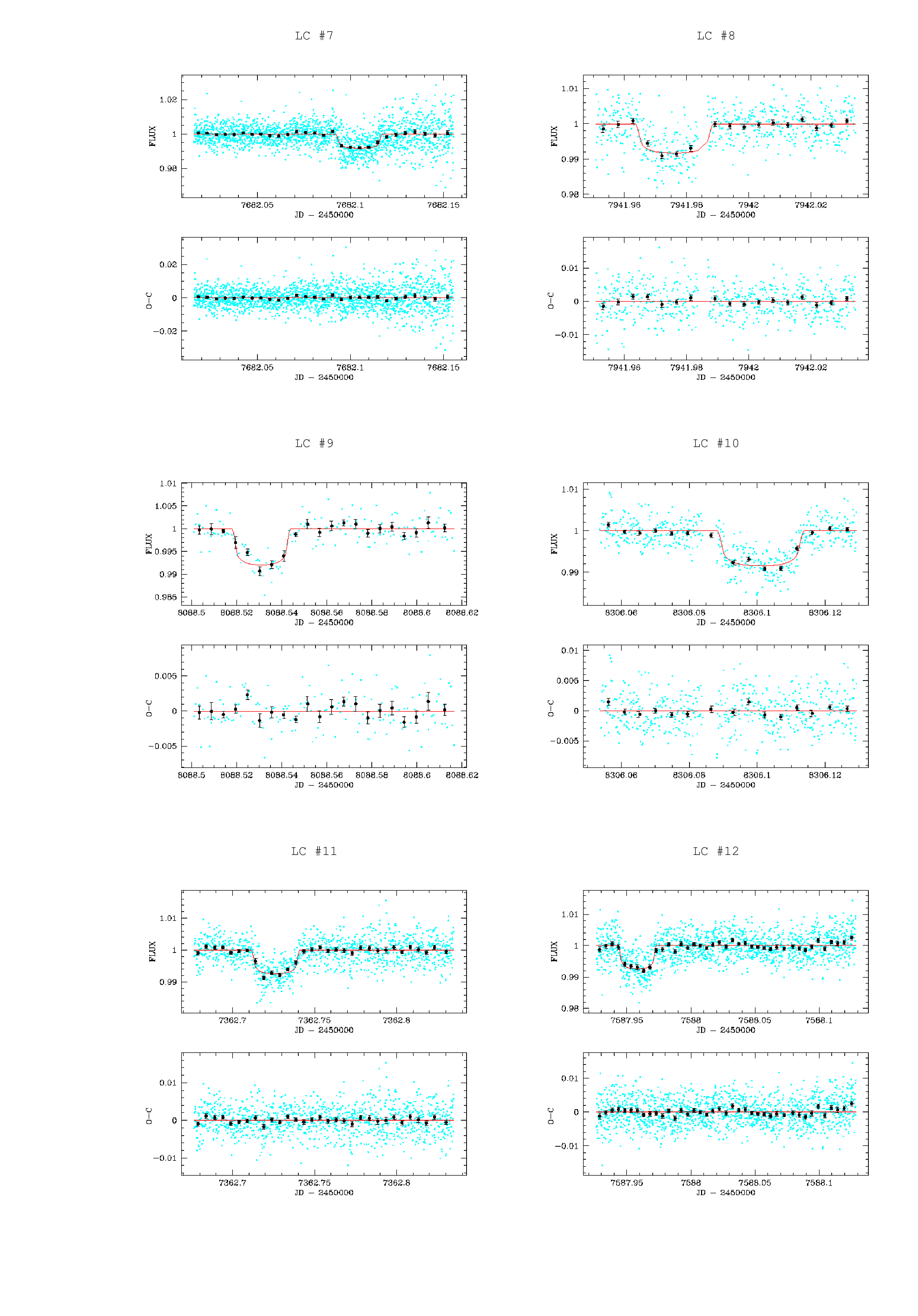}
    \caption{light curves (LC) \#7-12. See Tables \ref{tab:sets},\ref{tab:baselines}, and \ref{tab:ind_an} for the details.}
    \label{fig:plot2}
\end{figure}

\clearpage

\begin{figure}
    \centering
    \includegraphics[width=0.95\textwidth]{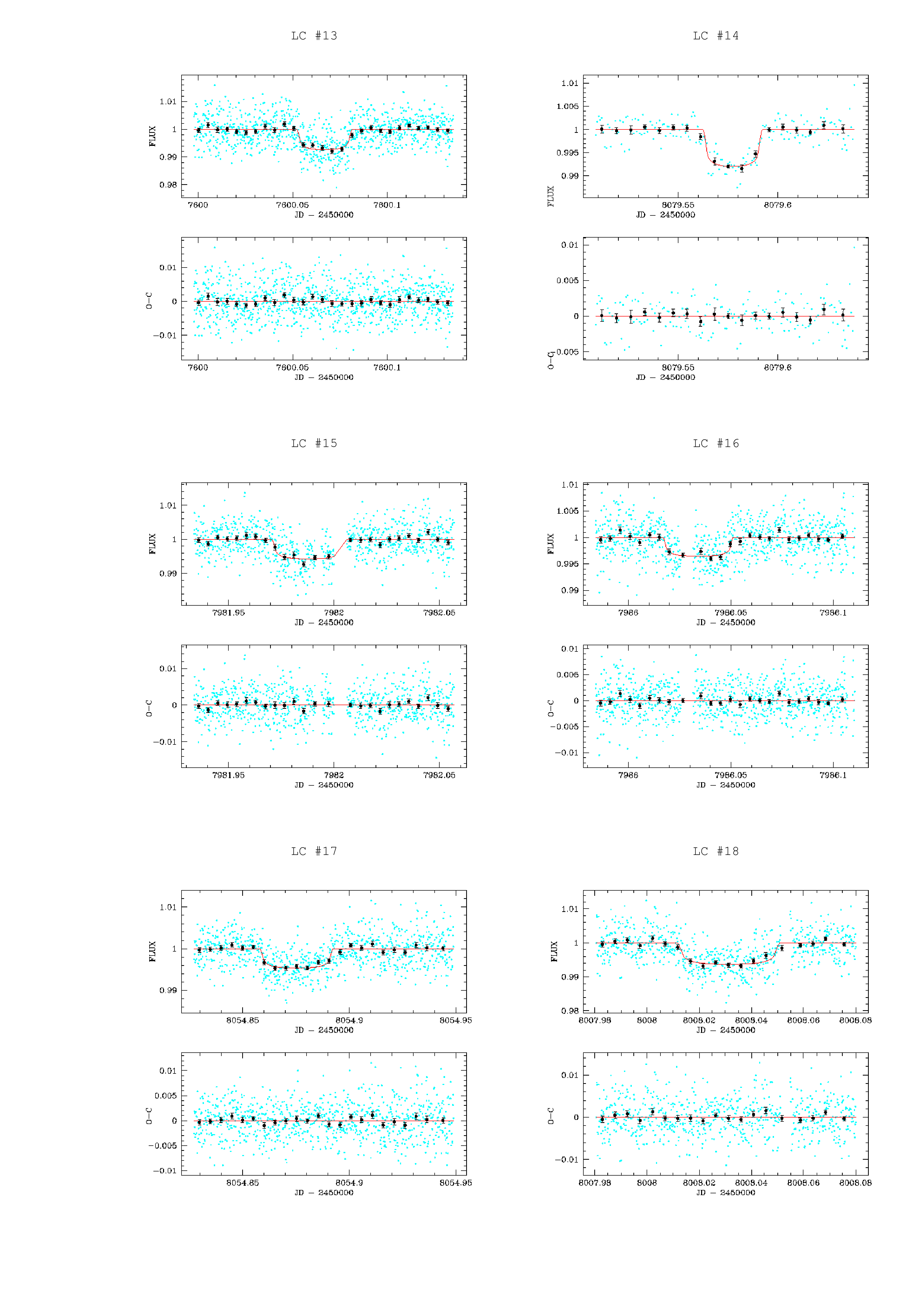}
    \caption{light curves (LC) \#13-18. See Tables \ref{tab:sets},\ref{tab:baselines}, and \ref{tab:ind_an} for the details.}
    \label{fig:plot3}
\end{figure}

\clearpage

\begin{figure}
    \centering
    \includegraphics[width=0.95\textwidth]{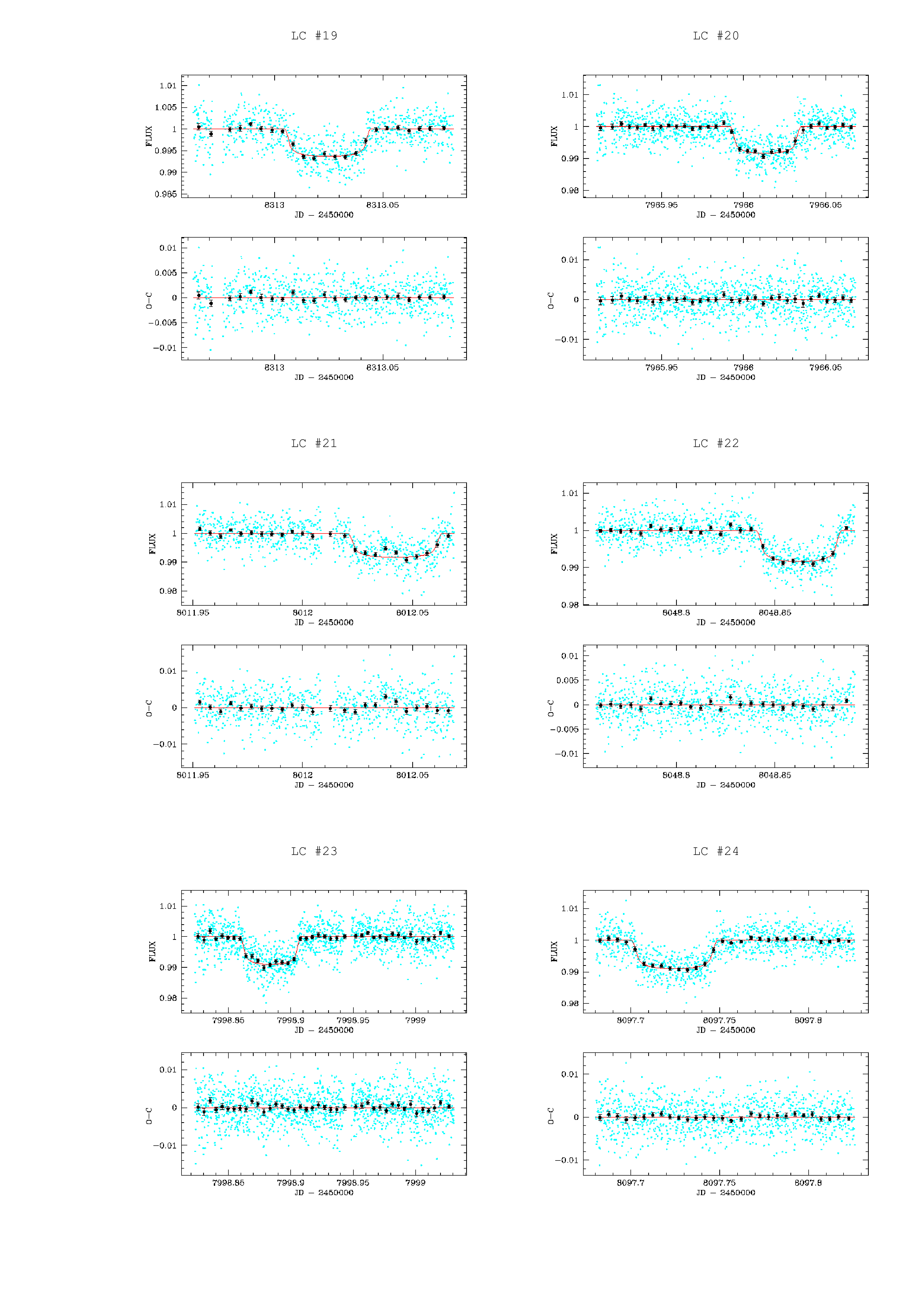}
    \caption{light curves (LC) \#19-24. See Tables \ref{tab:sets},\ref{tab:baselines}, and \ref{tab:ind_an} for the details.}
    \label{fig:plot4}
\end{figure}

\clearpage

\begin{figure}
    \centering
    \includegraphics[width=0.95\textwidth]{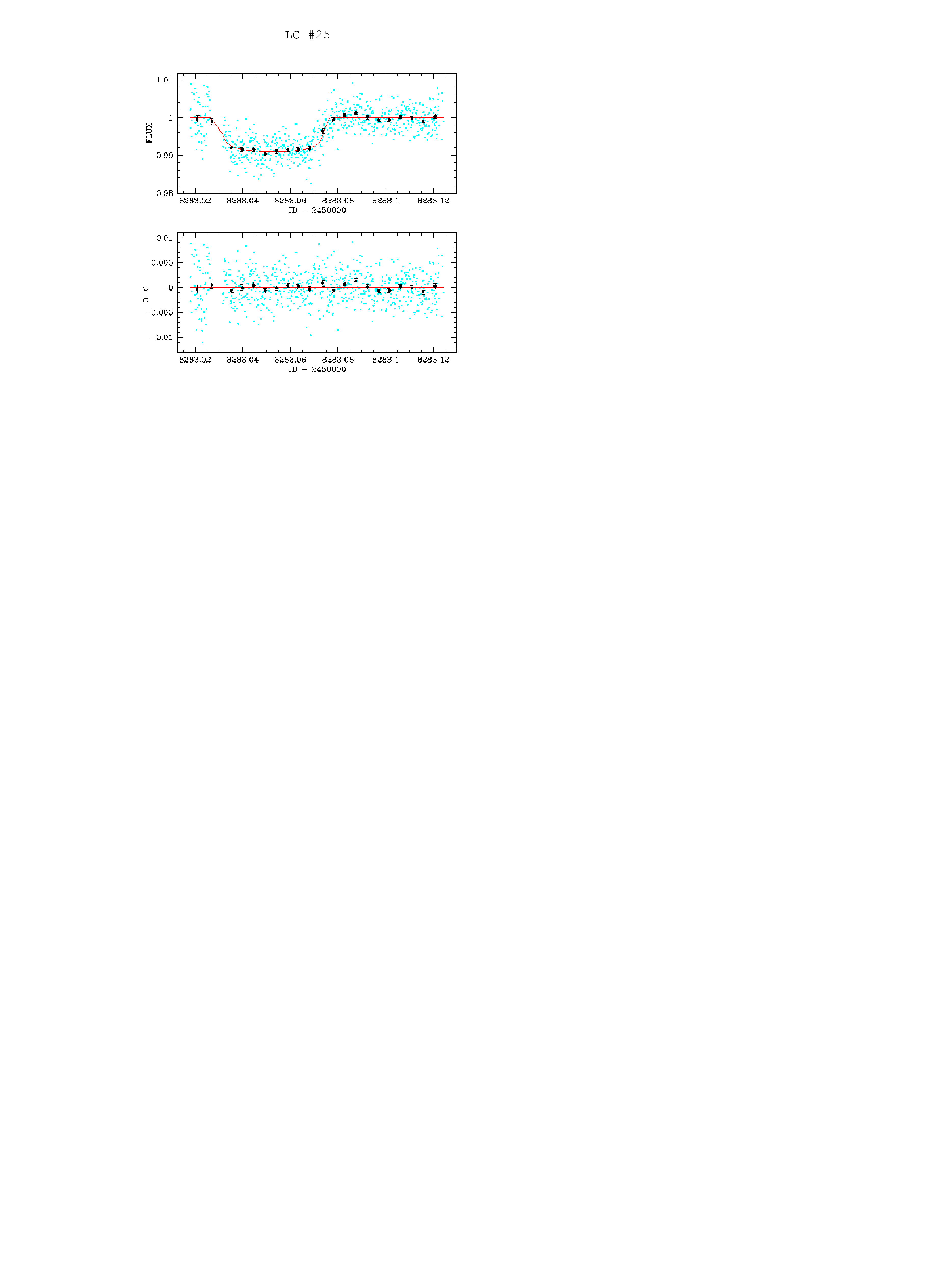}
    \caption{light curve (LC) \#25. See Tables \ref{tab:sets},\ref{tab:baselines}, and \ref{tab:ind_an} for the details.}
    \label{fig:plot5}
\end{figure}

\clearpage
%\newpage

%\section{Binned residuals RMS versus bin size plots}\label{B1}
%\clearpage

\begin{figure}
    \centering
    \includegraphics[width=1.0\textwidth]{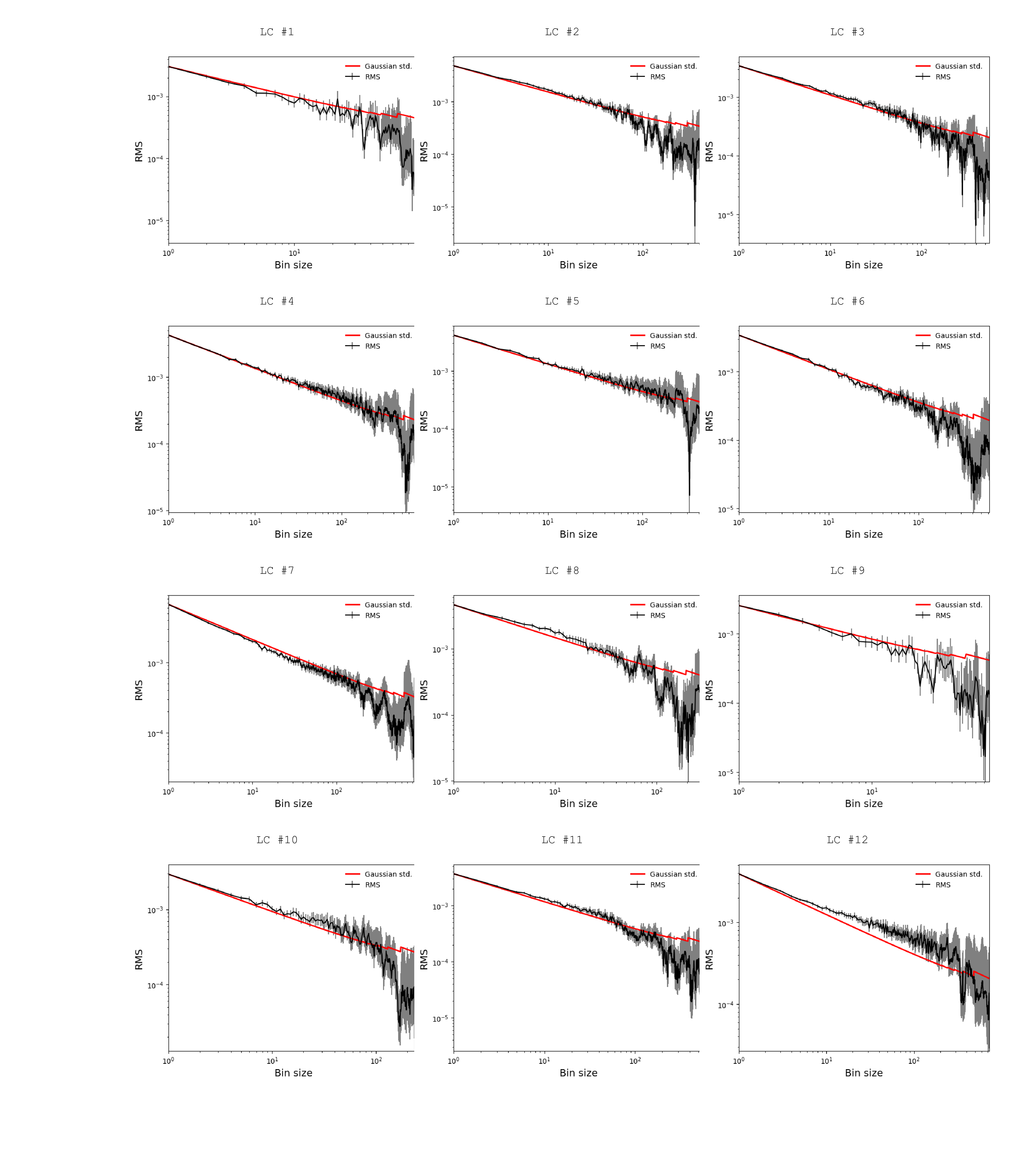}
    \caption{Binned residuals RMS versus bin size plots for light curves (LC) \#1-12.}
    \label{fig:plot_res}
\end{figure}

\begin{figure}
    \centering
    \includegraphics[width=1.0\textwidth]{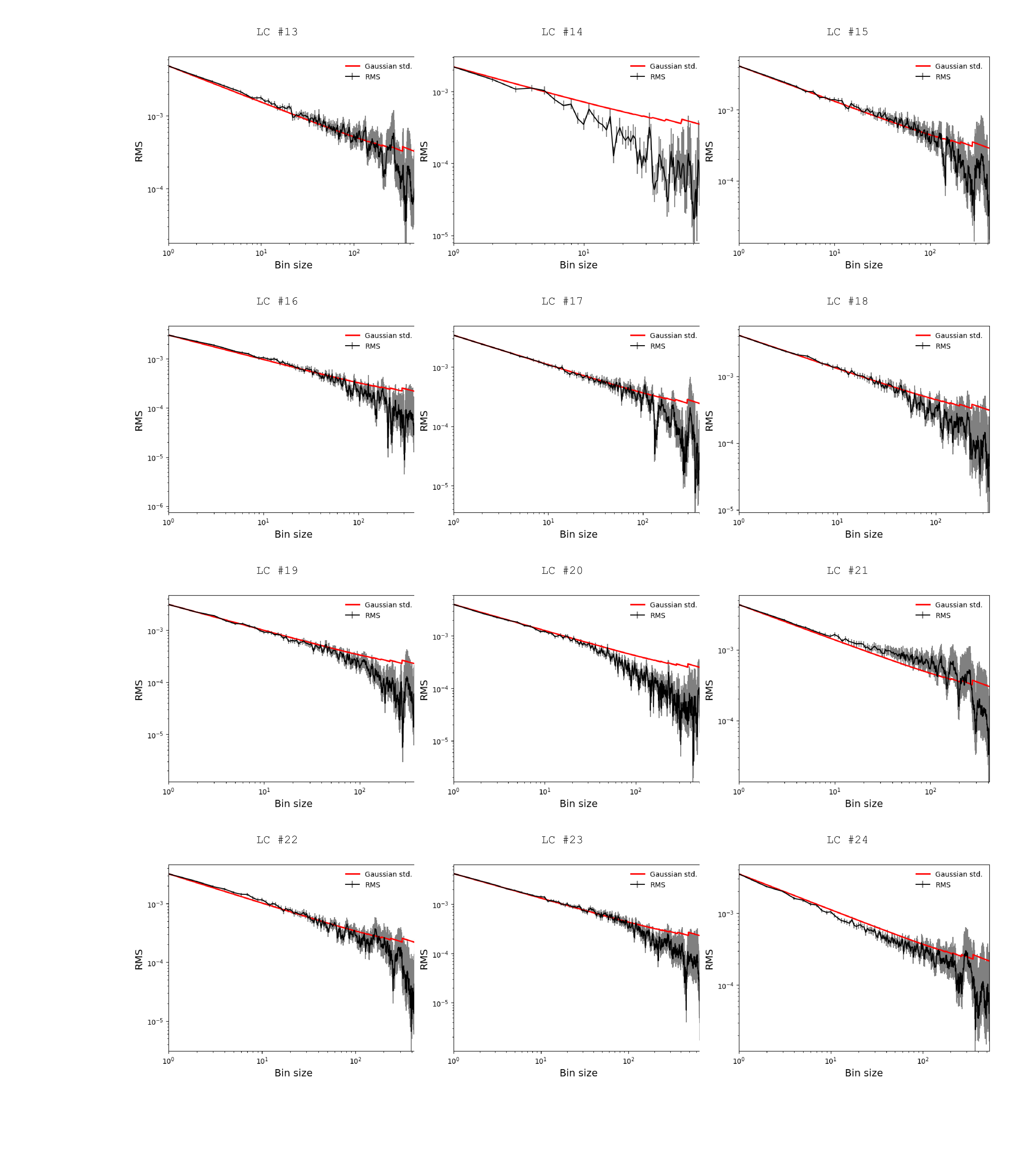}
    \caption{Binned residuals RMS versus bin size plots for light curves (LC) \#13-24.}
    \label{fig:plot2_res}
\end{figure}

\begin{figure}
    \centering
    \includegraphics[width=1.0\textwidth]{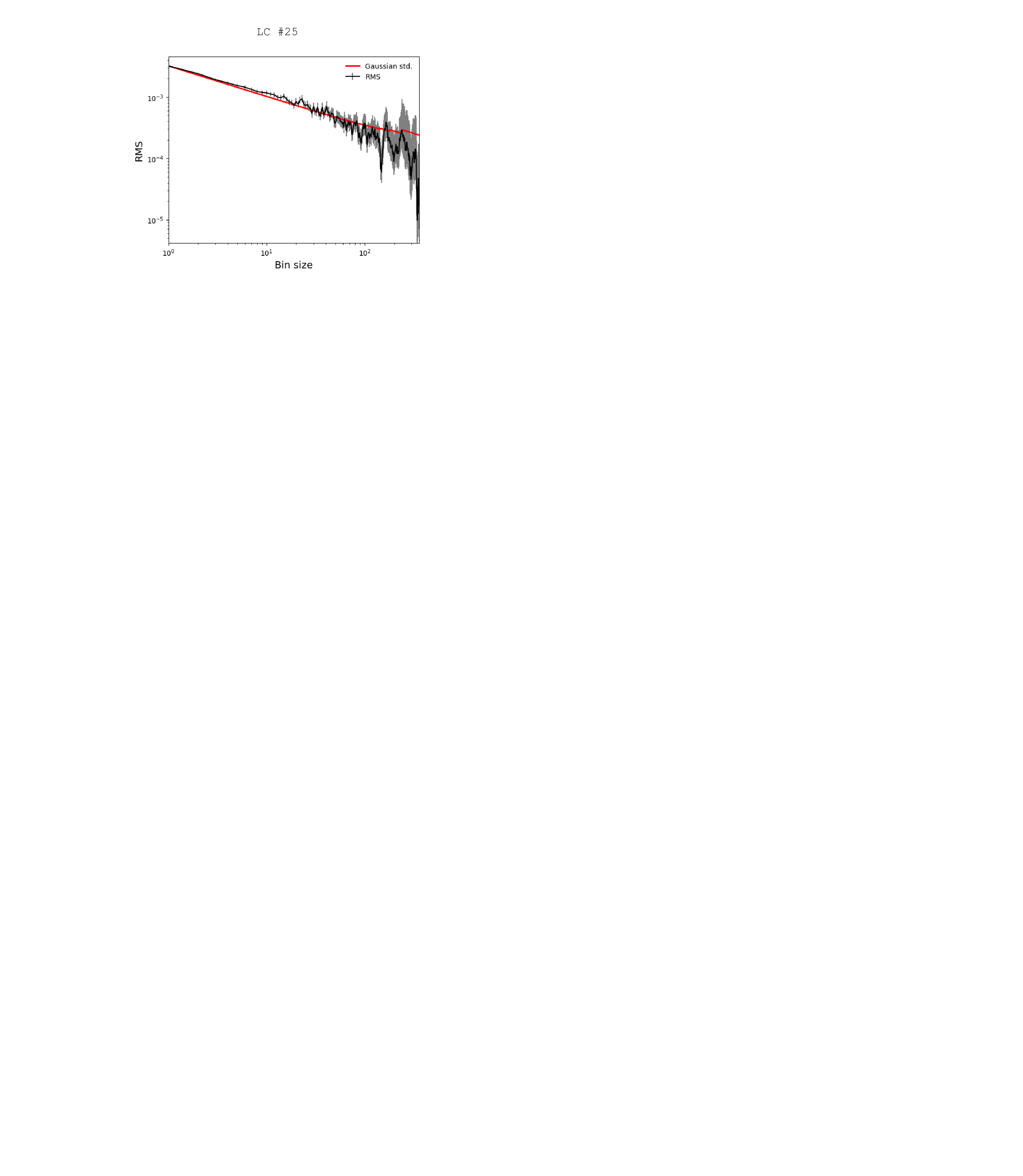}
    \caption{Binned residuals RMS versus bin size plots for light curves (LC) \#25.}
    \label{fig:plot3_res}
\end{figure}

\clearpage

%%%%%%%%%%%%%%%%%%%%%%%%%%%%%%%%%%%%%%%%%%%%%%%%%%

% Don't change these lines
\bsp	% typesetting comment
\label{lastpage}
\end{document}